# Student-Powered Digital Scholarship CoLab Project in the HKUST Library: Develop a Chinese Named-Entity Recognition (NER) Tool within One Semester from the Ground Up

**Sherry S.L. Yip**
Student from BSc in Data Science and Technology
Hong Kong University of Science and Technology, Hong Kong.
slyipae@connect.ust.hk

**Berry L. Han**
Student from BSc in Data Science and Technology
Hong Kong University of Science and Technology, Hong Kong.
lhanam@connect.ust.hk

**Holly H.Y. Chan**
Assistant Manager (Digital Humanities)
HKUST Library, Hong Kong University of Science and Technology, Hong Kong.
lbholly@ust.hk

## Abstract

*Starting in February 2024, the HKUST Library further extended the scope of AI literacy to AI utilization, which focuses on fostering student involvement in utilizing state-of-the-art technologies in the projects that initiated by the Library – named "Digital Scholarship (DS) CoLab". A key focus of the DS CoLab scheme has been on cultivating talents and enabling students to utilize advanced technologies in practical context. It aims to reinforce the library's role as a catalyst and hub for fostering multidisciplinary collaboration and cultivate the "can do spirit" among university members. The Library offers 1-2 projects per year for students to engage with advanced technologies in practical contexts while supporting the Library in tackling challenges and streamlining operational tasks. The tool that introduced in this paper was mainly developed by two of the authors, Sherry Yip Sau Lai and Berry Han Liuruo, as part-time student helpers under one of our DS CoLab scheme in the 2024 Spring Semester (February to May 2024). This paper details the complete journey from ideation to implementation of developing a Chinese Named-Entity Recognition (NER) Tool from the group up within one semester, from the initial research and planning stages to execution and come up a viable product. The collaborative spirit fostered by this project, with students playing a central role, exemplifies the power and potential of innovative educational models that prioritize hands-on learning with student involvement.*

# Table of Contents

## 1 Introduction

Starting in February 2024, the HKUST Library further extended the scope of AI literacy, such as through library workshop, to AI utilization, with the goal of fostering student involvement in utilizing state-of-the-art technologies in practical scenarios through the projects that initiated by the Library – namely "DS CoLab" (https://library.hkust.edu.hk/ds/ds-colab/) under the Library's new Digital Scholarship initiative. Under this framework, the Library offers student helper positions for undergraduate students to engage with advanced technologies in practical contexts while supporting the Library in tackling challenges and streamlining operational tasks in every Fall and Spring semester. This not only allows students to put their skills into practice but also benefits the Library the fresh perspectives and innovative ideas from students, creating a dynamic environment where learning and growth are mutual. The Library also proactively seeks opportunities for students to present their output from DS CoLab and helps them to connect with the academic world during their undergraduate years.

The Chinese Named-Entity Recognition (NER) Tool (see Table 1) that detailed in this paper was developed by two of the authors, Sherry Yip Sau Lai and Berry Han Liuruo, as part-time student helpers under the DS CoLab scheme in the 2024 Spring Semester during February to May 2024. Section 3 to Section 6 of this paper details the complete journey from ideation to implementation of developing the tool from the group up within a 4-month timeframe in one semester. A key focus of this project has been on cultivating talent, enabling students to utilize advanced technologies in practical context, giving them high flexibility and involvement in the entire project lifecycle, from the initial research and planning stages to execution and come up a viable product. Through project-based work, students benefit from a dual approach of mentorship from library staff and self-directed exploration. The collaborative spirit fostered by this project, with students playing a central role, demonstrated the power and potential of the model that could be provided by academic libraries which is to prioritize hands-on learning with student involvement, reinforces the Library's role as a catalyst and hub for fostering multidisciplinary collaboration and cultivate the "can do spirit" among university members.

**Table 1**

*Project Details and Featured Links*

| Featured | URL |
|---|---|
| Project page | https://library.hkust.edu.hk/ds/project/p001/ |
| Students sharing | https://library.hkust.edu.hk/ds/student-sharing/student-learning-journey-han-liuruo-berry-yip-sau-lai-sherry-ds-p001/ |
| GitHub repository | https://github.com/hkust-lib-ds/P001-PUBLIC_Chinese-NER-Tool |
| Manual guide | https://github.com/hkust-lib-ds/P001-PUBLIC_Chinese-NER-Tool/blob/main/manual.md |

## 2 Literature Review

In the evolving landscape of higher education, academic libraries play a vital role in supporting student engagement and scholarship, preparing students for success in the 21st century (Miller et al., 2019; McCarl, 2021; Holthe & See, 2022). Academic libraries are transforming from mere information access points to a dynamic hub where students can create, connect, and grow by fostering an environment of experimentation and collaboration within the library, including but not limited to providing opportunities for staff and students to exchange their knowledge with one another and encouraging them to experiment with new technologies (McGinn & Coats, 2023). Experiential learning in academic libraries has been increasingly recognized for its potential to enhance student engagement and learning, particularly through digital scholarship and digital humanities projects (Natal & Remaklus, 2023). Students can be involved in project-based learning (PBL) from conception to completion, and to gain unique learning experience and develop a range of transferrable skills (Fede et al., 2017; McCarl, 2021; Natal & Remaklus, 2023), which in turn enhances the library's role as a learning hub on campus.

Many libraries and labs in the universities worldwide have already adopted the concept of experiential learning and project-based learning, actively engaging students in hands-on activities and collaborative projects to enrich their learning experiences during their time at university. For example, the Gettysburg College's Musselman Library launched a student-focused, library-led, ten-week paid summer program of Digital Scholarship Summer Fellowship for letting undergraduates expose to a range of digital tools with appropriate partners about research practices and possibilities, and successfully developed several public-facing student research (Wertzberger & Miessler, 2017). The National University of Singapore Libraries offers Undergraduate Research Library Fellowship (https://nus.edu.sg/nuslibraries/services-help/fellowships-grants) for students to pursue research projects with library collection, supported by mentorship from library staff. The Digital Humanities Lab of the University of Exeter provides internship opportunities for undergraduate students (https://digitalhumanities.exeter.ac.uk/category/interns/) to participate in various projects and gain essential practical skills in different digital techniques. The survey conducted by Fede et al. (2017) also showed that the growth of students' transferable skills, such as communication and problem-solving, as well as civic behaviors and attitudes, can be fostered through the job opportunities provided by universities. Results showed that creating meaningful work experiences in these positions mirrors the benefits of other experiential education opportunities. It highlighted the potential of university employment as a pathway for skill enhancement and emphasizing partnerships in these roles could provide mutual benefits for universities, students, and the cooperative partner, such as the library in our case.

In light of the insights inspired above, the framework of the HKUST Library's DS CoLab was developed as detailed in the next section, which focuses on providing student helper positions for undergraduates, allowing them to gain valuable project-based experiential learning experience to engage with digital tools and cutting-edge technologies in real-world contexts.

## 3 Project Cycle

The core of project-based learning (PBL) is that the project must be meaningful and doable, with an emphasis on incorporating student voice and choice into the project, as well as setting a clear project timeline (Markham, 2011). As stated by Holthe and See (2022), effective partnerships rely on continual yet constructive communication. When working with individual such as students, we should connect with them regularly, share updates and progress reports, ask thought-provoking questions, and celebrate collective achievements. PBL is not merely doing projects, but instead must involve formative feedback, guidelines, systematic evaluation, and a "shift in power" to the student (Markham, 2011).

Figure 1 illustrates the design of the HKUST Library's DS CoLab framework. To start, the Library carries out a needs-based analysis to determine the project topic, establish clear goals, and assess feasibility. Next, it moves to student recruitment, targeting individuals whose skills and interests match the project's requirements. The students are recruited as part-time student helpers during the Fall or Spring semester for an approximately four-month period, with an expectation of dedicating 10 hours per week for the project. Once the project is underway, students are involved throughout the entire process with high flexibility of involvement in every stage of the project. They can choose to work anywhere at any time, yet they are required to meet with library staff, who serve as mentors and/or advisors, every week to discuss and report progress regularly. At the end of their employment period, students are asked to write about their project experiences, detailing what they did, what things they newly learnt, the challenges they encountered, and how they overcame them. The Library also actively seeks opportunities for students to showcase their work from the DS CoLab and facilitates connections with the academic community. If students are interested, they can collaborate with library staff mentors again for additional scholarly and promotional activities on a voluntary basis.

**Figure 1**

*Cycle of a DS CoLab project*

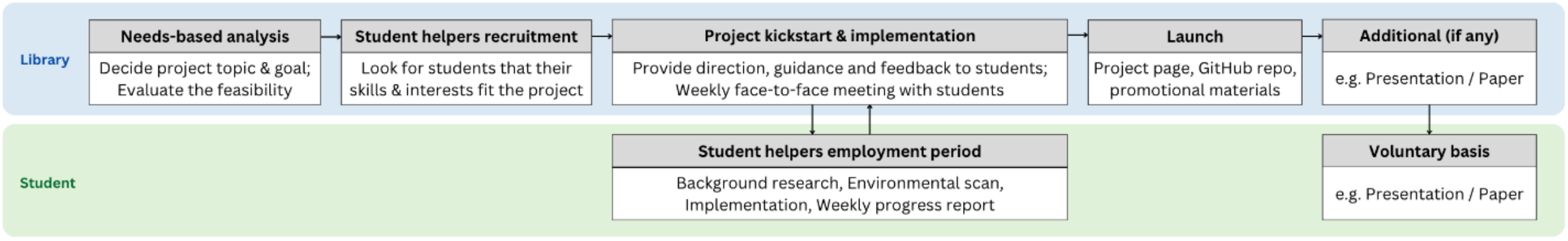

The Chinese Named-Entity Recognition (NER) Tool featured in this paper is one of the first projects as we embark on this DS CoLab framework. When working on the HKUST Digital Humanities Initiative's projects (https://digitalhumanities.hkust.edu.hk/), it has come to our attention that there are various models and tools available online for doing NER tasks but many of them need coding and time for customization in order to reach the desired results. The technical complexity of such methods may be an obstacle for those researchers without strong coding skills.

Therefore, we set the goal of this project as to develop a user-friendly tool that can streamline the Chinese NER process, allowing non-technical researchers to overcome the technical barriers and to more conveniently utilize NER in their work, particularly for the Chinese language research community. Additionally, the tool can also be used for future digital humanities and library projects, enhancing our efficiency in similar tasks and enabling non-technical staff to perform these tasks as well. After confirming our objectives, several tools and models are shortlisted, together with our expectation, shared with two of our authors Sherry and Berry. They then carried out comprehensive evaluation on the shortlisted items, with research on more additional tools and models as part of environmental assessment. Afterward, timeline and workplan were discussed and clearly outlined to ensure that a viable product can be delivered within a four-month period. Considerable flexibility was given for students in terms of user interface and function design, meanwhile, a range of digital tools were introduced to them during the development phase. Weekly meetings were conducted to review the progress and offer guidance. Librarians from different sections were also involved to provide user perspective insights to make our tool more user-friendly and engaging. The following sections will detail the environmental scan findings, the functionality design and the challenges encountered, revealing the story behind the creation of this tool.

## 4 Comparisons of Existing NER Tools and Models

Named-Entity Recognition (NER) is a natural language processing technique that can automatically identify and categorize key elements such as people, organizations, locations, dates, and other important concepts within a large amount of text. This technique enables researchers to conduct insightful analysis of textual information relevant to their research efficiently.

Before developing our own NER tool, it is essential to examine existing products and models in the field. The environmental scan examination can help identify both the strengths and shortcomings of current solutions, revealing gaps that our tool could fill. By analyzing a range of NER tools, we can determine which features are the most effective and applicable in various contexts. This comparative analysis guides our selection of methods and technologies, ensuring that the tool that is going to develop to the later stage would be able to incorporate the best practices while addressing unmet needs.

### 4.1 Environmental Scan on Existing NER Tools

**Table 2**

*Comparison of Existing NER Tools for Chinese Corpus*

| Tool | Targeted Languages | Entity Classes | Deliver Format | Main Features | Operational Functionality |
|---|---|---|---|---|---|
| **WYD** | Simple Chinese<br>English<br>Yongtai (永泰文书) | 5 | Website<br>Cloud-driven | Structure Management<br>NER Annotation<br>Relationship Annotation<br>Graph Generation | File Upload & Text Inputting<br>Paragraph Segmentation<br>Automatic Word Split<br>Customized Definition Upload<br>Automatic Recognition<br>(Word Split, NER, Relationship, Graph)<br>Manually Correction<br>(Word Split, NER, Relationship)<br>Display Filtering<br>(NER, Relationship)<br>Data Export (CSV, JSON, et.al.) |
| **CORPRO** | Simple Chinese<br>Traditional Chinese | 0 | Software | Word Split<br>Vocabulary Frequency<br>Keyword in Context<br>Keyword Dispersion<br>Word Trend<br>N-Grams<br>Word Clusters<br>Collocation | Corpus Management<br>Automatic Word Split<br>Manual Word Split Correction<br>Word Category Management<br>Dictionary Editing<br>(Import, Stop Words Definition)<br>Corpus Category Definition<br>Data Export |
| **MARKUS** | Simple Chinese<br>Traditional Chinese<br>Korean | 9 Classes<br>7 Packages | Website<br>Online Storage | NER Annotation<br>Online reference tools | File Upload & Text Inputting<br>Customized Definition Upload<br>Automatic / Manually Tagging<br>Display Filtering<br>Data Export |

**WYD (**Research Center for Digital Humanities of Peking University, 2023, **https://wyd.pkudh.net/)**, developed by the Digital Humanities Research Center of Peking University, is an intelligent annotation cloud platform. Users can store their data as projects online. There are four main features offered: structure management, name entity annotation, relationship annotation, and graph generation. Corpuses, imported by file uploading or text inputting, can be first pre-processed with paragraph segmentation and word splitting in "*Structure*" section. Then name entities are annotated in "*Entity*" session and relationship among those entities can be recognized in "*Relationship*" session. Automatic recognition, manual correction, and customized definition uploading are allowed to conveniently define entities and relationships. Corresponding data of entities and relationships can be downloaded as single or multiple files in CSV or JSON format. In "G*raph*" session, a relation network will be created based on the defined relationships and entities, which can be exported in various formats for convenient corporation with common graph analytic tools.

WYD provides a model sample of designing an interactive user interface, which serves as inspiration and reference for us to develop the operational functionalities of our tool.

**CORPOR (**闕河嘉&陳光華, 2016, **https://nlp.cse.ntou.edu.tw/CORPRO/)** is a software-based analytic tool developed by Que and Chen from National Taiwan University, which provides Chinese Word splitting and plenty of search and visualization functionalities. It allows users to use the data on their local computer, which means it can manage large datasets as long as the users' computer has enough disk space and sufficient processing power. It also allows users to group and label files for easier management. Although the tool does not support name entity recognition, as they focus on word splitting, some of the visualization methods are applicable for name entity. For example, given one or several specified keywords, the chart for keyword dispersion shows the appearance position, and the graph for word trend shows the variation of frequency. Apart from visualization, there are many useful functions for neighbourhood analysis, such as keyword-in-context search which search for corpus and correspond neighbouring content, N-gram word search which analyses consecutive combination before and after a centre word, and collaboration of word cluster which analyses words that appear within a specified range given a targeted word. The tools also support various operations on vocabulary, including definition uploading and manual correction of word splitting, word stopping, and word category defining which allows customized grouping words to analysing as a whole.

CORPOR proves instrumental in facilitating comprehensive professional analysis through a diverse array of operations and functions; however, its complexity may present challenges for non-technical users seeking rapid proficiency. Given our objective of creating a user-centric tool, we selectively take reference on some of the visualization and operational design elements from their product rather than their architecture.

**MARKUS (**Ho & Weerdt, 2014, **https://github.com/dHumanities/markus)** is a website with online record storage for account users similar to WYD. Most of the functionalities and operations in the "*Entity*" part of WYD are supported in MARKUS, such as automatic and manual entity annotation, annotation of definition uploading and display filtering. However, different annotation methods correspond to different entries in MARKUS, affecting the usage flexibility.

After comparing with WYD, we decided to mainly follow the design of WYD instead of this tool because the former one includes more comprehensive functionalities and has a more user-friendly interface. Though this tool incorporates several online reference tools to search geographical biographical knowledge which is very useful feature, considering our time limitations and the absence of a necessary database, we ultimately chose not to implement this feature in our tool.

After comparing the above three platforms, we selected a list of features as outlined in Table 3 for our tool.

**Table 3**

*List of Selected Features for Our Tool*

| Feature | Reference / Inspiration |
|---|---|
| **Corpus Import** | Similar to WYD and MARKUS, both file uploading and text inputting are used for Corpus importing. Text importing is convenient for trial and the case for single file, while file uploading can manage multiple files more efficiently. |
| **Automatic and Manual Entity Tagging** | As the main functionality of our tool, automatic name entity tagging provides a powerful way to recognize entities in the text. However, mistakes are almost unavoidable. Therefore, similar to all the three existing tools - WYD, CORPRO and MARKUS, our tool also supports manual correction, including adding, deleting and editing of entities. |
| **Customized Definition Uploading** | Manual correction of name entity is not sufficient to address the shortage of automatic tagging, because of the inefficiency of repetitive operations, especially when there are considerate number of texts or mistakes. Uploading a file of customized definition of entities is an effective solution, which is also practiced in WYD and MARKUS. Additionally, this functionality also enhances flexibility and sustainability, as it allows leveraging other or more enhanced recognition models in the future for entity definition. |
| **Different Classification Methods and Display Filtering of Entity Annotation** | Inheriting from WYD and MARKUS, entities are originally classified by entity class. The colours of annotation are various across entities classes. Considering that entities from different classes may worth studying as a group and being inspired by the word category definition in CORPOR, we developed an additional type of classification, the entity group, which allows users to freely collect a few entities. Additionally, since it is common that several entities may refer to a semantically identical concept (e.g. a character with alternative names), we include "Entity Alias" in our tool for customized combining equivalent entities. By choosing to apply or not apply the defined Entity Alias, users can specify to whether treat the entities in an Entity Alias as a whole or separately. Display filtering is supported in the unit of these three types of classification (i.e. entity class, entity group, and entity alias), which means display annotation and data result of a certain subset of entities. |
| **Various Data Visualizations** | To better assist data analysis, we visualize the data of entities from different perspectives using four kinds of charts, including: (1) colour boxes for entity class-instance overview, which summarize the number of entity instances of each entity class; (2) bar chart for frequency of entity instances, which shows the number of occurrence of the selected entities; (3) scatter plot for position of entity instances, which shows the positions that the selected entities appeared in the text; (4) line chart for cross-document frequency of entity instances, which shows the variation trend of frequency across documents for a selected entity. The last two types of charts are inspired by the visualization of keyword dispersion and word trend in CORPRO. Particularly, the targeted entities of the charts (except the first one, colour boxes for entity-class overview) correspond to the display filtering. |
| **Data Export** | For management convenience, useful data for a corpus are packed as a zip file for downloading. |

## 4.2 Environmental Scan on NER Models

Evaluation of commonly used Named-Entity Recognition (NER) models is conducted to identify the most suitable option for our tool. As we targeted at Hong Kong users as our first priority, the model's performance on Traditional Chinese text was a critical consideration. Additionally, factors such as API usage limitations, model accuracy, and flexibility for fine-tuning were also essential in our selection process. After thorough research and comparative analysis of the available models on the market, we focused on four widely adopted NER models: HanLP (He & Choi, 2021), PaddleNLP (PaddleNLP Contributors, 2021), THULAC (Sun et al., 2016), and CKIP (Academia

Sinica, 2023). Table 4 below provides a detailed comparison of these models based on their strengths, limitations, and suitability.

**Table 4**

*Comparison of NER Models*

| Model | Update Time | Supported Languages | Supported Functions | Default Entity Classes | Shortcomings |
|---|---|---|---|---|---|
| **HanLP** | 2021 | Simple Chinese, Traditional Chinese | Named entity recognition, word segmentation, POS tagging, dependency parsing, sentiment analysis, etc. | 18 or 30 | Only allows 4 API requests per minute for basic users |
| **PaddleNLP** | 2021 | Simple Chinese | Named entity recognition, word segmentation, POS tagging, dependency parsing, sentiment analysis, etc. | 66 | Performs poorly on Traditional Chinese text |
| **THULAC** | 2017 | Simple Chinese | Word segmentation, POS tagging | | Lacks named entity recognition functionality |
| **CKIP** | 2021 | Simple Chinese, Traditional Chinese | Named entity recognition, word segmentation, POS tagging | 18 | Training data is limited |

**HanLP model** has been trained with rich data and provides relatively comprehensive functions. However, it only allows four API requests per minute for new basic users. This limits the efficiency of text processing, so we opted not to select this model for our tool.

**PaddleNLP model** is highly efficient for processing, but it was found to be inaccurate in recognizing Traditional Chinese text after multiple trials. Considering that our tool targets at non-technical users based in Hong Kong, the ability to handle Traditional Chinese is important. Therefore, we opted not to select this model for our tool.

**THULAC model** was developed by the Tsinghua University and offers word segmentation and POS tagging functions. However, it does not offer Named-Entity Recognition (NER) capability that our tool was seeking. Additionally, this model was trained several years ago, which there are more other models rolled out in the rapidly changing AI field. Consequently, we opted not to select this model for our tool.

**CKIP model** has no restrictions on the usage limits and the code is fully open-source, allowing users to optimize the model according to their needs, despite its relatively limited training data compared to HanLP and PaddleNLP. According to the official documentation, its language model is trained on the ZhWiki and CNA datasets, the word segmentation (WS) and POS tagging models are trained on the ASBC dataset, and the NER model is trained on the OntoNotes dataset, which proves the model's reliability and authority. Additionally, unlike other models, the CKIP model was trained entirely on Traditional Chinese data, making it more reliable and accurate for recognizing Traditional Chinese text.

In conclusion, despite CKIP's smaller training dataset and fewer features compared to HanLP and PaddleNLP, its strong support for Traditional Chinese, open-source code, unlimited usage, and model fine-tuning capabilities makes it the best fit for our tool. Given the focus on serving Hong Kong users, CKIP's accuracy and reliability in handling Traditional Chinese text outweigh the advantages of other models, and therefore, we decided to select this model as the foundation of our tool for the Chinese NER task of annotation.

## 5 Features and Applications of Our Tool

### 5.1 Terminologies

Four important terminologies are used in our tool, the description and examples of which are summarized in Table 5.

**Entity Instance (實體實例)** refers to the words or short phrases that are recognized as entity, which are annotated in our tool. Some others commonly used terms for the same concept are entity, name entity, entity occurrence, entity token.

**Entity Class (實體類別)** refers to the category an entity instance that can belong to. In our tool, each entity instance belongs to one entity class. Other commonly used terms for this concept are entity tag and NER tag.

**Entity Group (實體群組)** refers to a collection of entity instances defined by users, which can include any entity instances in any entity class. This concept is primarily designed for the convenience of studying a certain subset of entity instances.

**Entity Alias (實體別名)** is used to define a collection of entity instances that refer to the same concept. The primary targeted scenario of this term is that characters in literature usually have several alternative names, which are extendable to other types of entity such as location.

**Table 5**

*Four Important Terminologies in Our Tool*

| Terminology | Description | Example |
|---|---|---|
| Entity Instance (實體實例) | The word in the corpus that is recognized as entity | 唐三藏(Tang Sanzang, Tripitaka, or Tang Monk)<br>孫悟空(Sun Wukong, or Monkey King)<br>豬八戒(Zhu Bajie, or Pigsy)<br><br>金箍棒 (Jingu Bang, or Golden Hoop)<br>筋斗雲 (Jindouyun, Nimbus Cloud, or Somersault Cloud) |
| | | 中國(China)<br>北京(Beijing) |
| Entity Class (實體類別) | The category that an entity instance can belong to. | 人名(Person)<br>產品(Product)<br>地理區(Location) |
| Entity Group (實體群組) | A customized defined collection of any different entity instances | An Entity Group named 【孫悟空打妖怪關鍵詞】may include:<br>孫悟空(Sun Wukong, or Monkey King)<br>金箍棒 (Jingu Bang, or Golden Hoop)<br>筋斗雲 (Jindouyun, Nimbus Cloud, or Somersault Cloud)<br>An Entity Group named 【道教人物角色】may include:<br>玉皇大帝(Yudi, or Jade Emperor)<br>菩提祖師 (Puti Zushi, or Patriarch Bodhi)<br>太上老君 (Taishang Laojun, or The Supreme Venerable Sovereign) |
| Entity Alias (實體別名) | A collection of entity instances referring to the same concept | An Entity Alias named 【孫悟空】 may include:<br>孫悟空(Sun Wukong, or Monkey King)<br>悟空(Wukong)<br>孫行者 (Sun Xingzhe, or Pilgrim Sun)<br>齊天大聖 (Qitian Dasheng, or Great Sage Equal to Heaven) |

## 5.2 Features of Our Tool

In this section, the features of our tool as outlined in Table 6 will be explained in detail, including: (1) Corpus import; (2) Auto Named-Entity Recognition (NER); (3) Data management of entity instance, class, group, and alias; (4) Data visualization; and (5) Data export.

Figure 2 below shows the layout of our tool. Most of the operations for managing entity instance, class, group and alias are on the sidebar in the left-hand side, while corpus management, display of annotated text and visualization are on the main area in the right-hand side. The overall flow of using our tool is depicted in Figure 3.

**Figure 2**

*Layout of Our Tool*

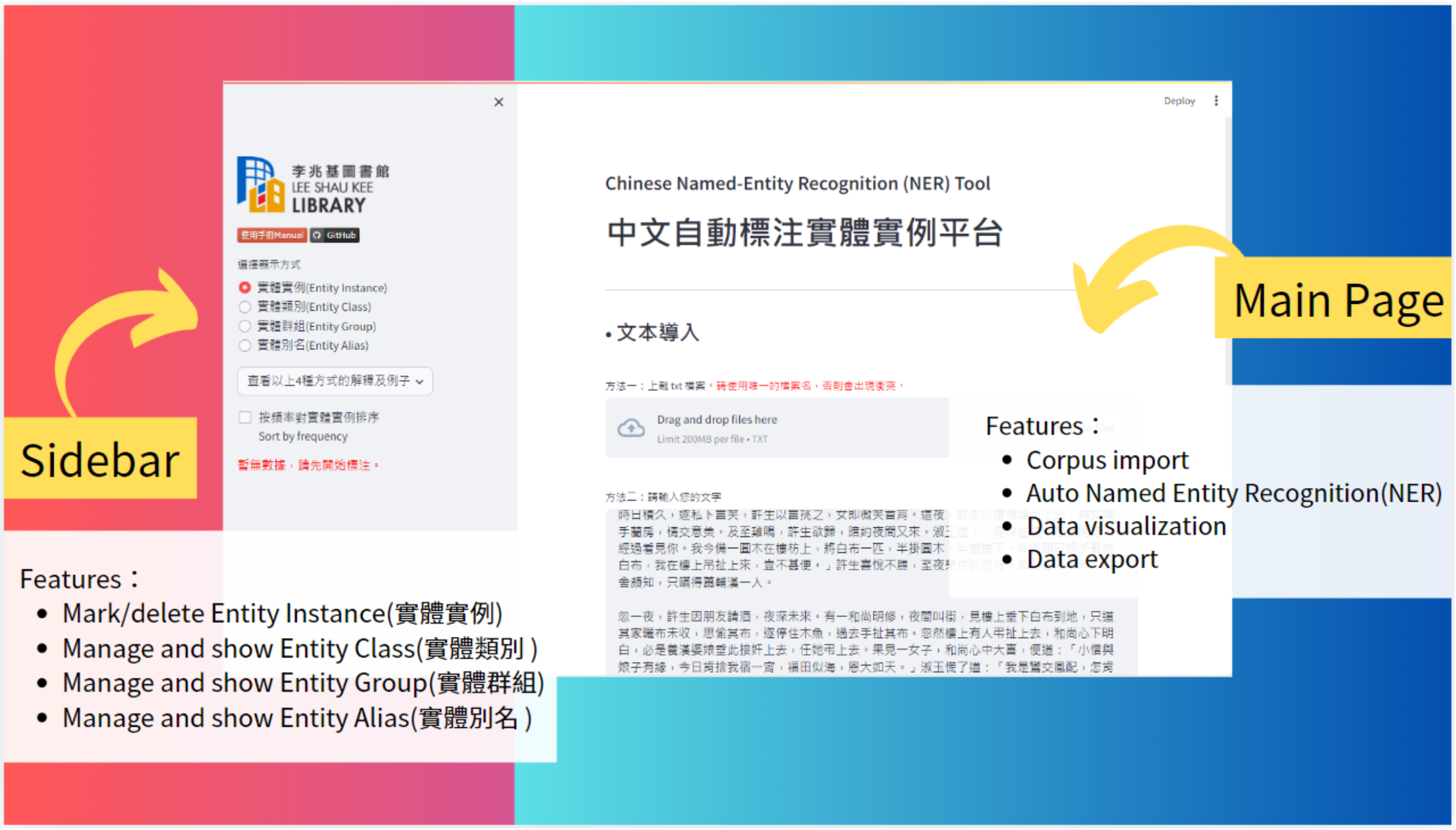

**Figure 3**

*Flow of Our Tool*

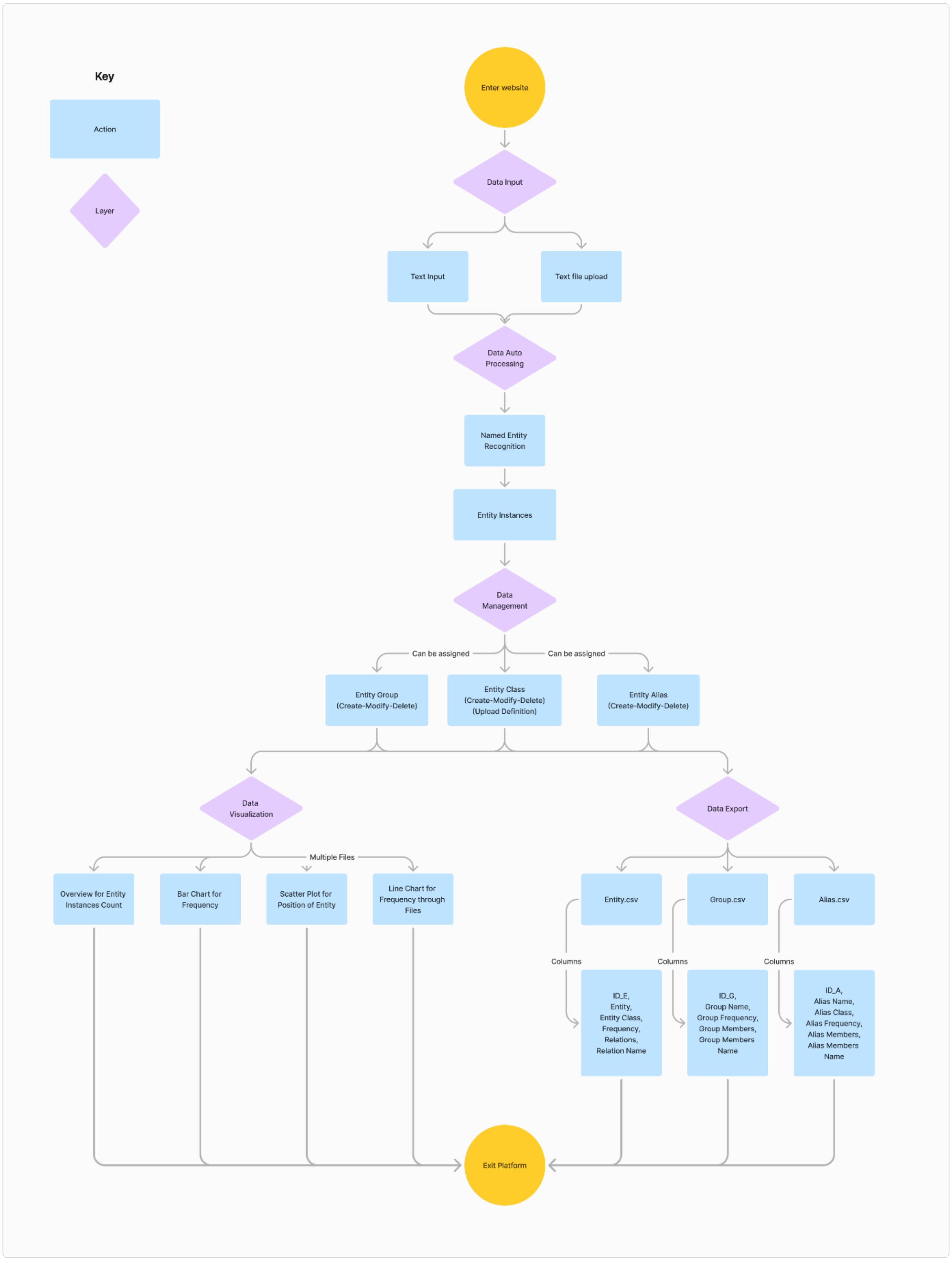

**Table 6**

*Summary of the Features in Our Tool*

| Feature | Details | Effect Scope |
|---|---|---|
| Corpus Import | Text input/txt file upload | |
| Auto Named Entity Recognition (NER) | Automatic recognition of entity instances using the BERT-base model from CKIP | All files |
| Customized Definition Upload | Upload a file with customized definition of entity class and entity instances | All files |
| Entity Instance Management | Manually add, edit or delete entity instances, or batch defined by definition upload | All files |
| Entity Class Management | Filter and display entity instances based on entity class<br>Manually add, edit or delete entity instances, or batch defined via definition upload | All files |
| Entity Group Management | Filter and display entity instances based on entity groups<br>Manually create, edit, and delete entity groups | Currently viewed file |
| Entity Alias Management | Filter and display entity instances based on aliases<br>Manually create, edit, and delete entity aliases | Currently viewed file |
| Data Visualization | Four types of visualizations to intuitively present the frequency and position of entity instances | Currently viewed file |
| Data Export | One-click download the zip file of three CSV files that summarized the information of entity instances, entity groups, and entity group | Currently viewed file |

## 5.3 Application of Our Tool – Scenario Example (Single File)

A manual guide is available on our GitHub repository (https://github.com/hkust-lib-ds/P001-PUBLIC_Chinese-NER-Tool/blob/main/manual.md), outlining the detailed steps for each function of our tool. In the following section, we will demonstrate the capabilities of our tool more in depth by using the first chapter of "*包公案—龍圖公案 (Bao Gong Case)*" (Full text from: http://open-lit.com/book.php?bid=189) as sample data to showcase the functions and analysis the users can perform using our tool.

### 5.3.1. Corpus import

To start with, paste the required text into the text input area as shown in Figure 4.

**Figure 4**

*Import Data via Inputting/Pasting Text in the Text Area*

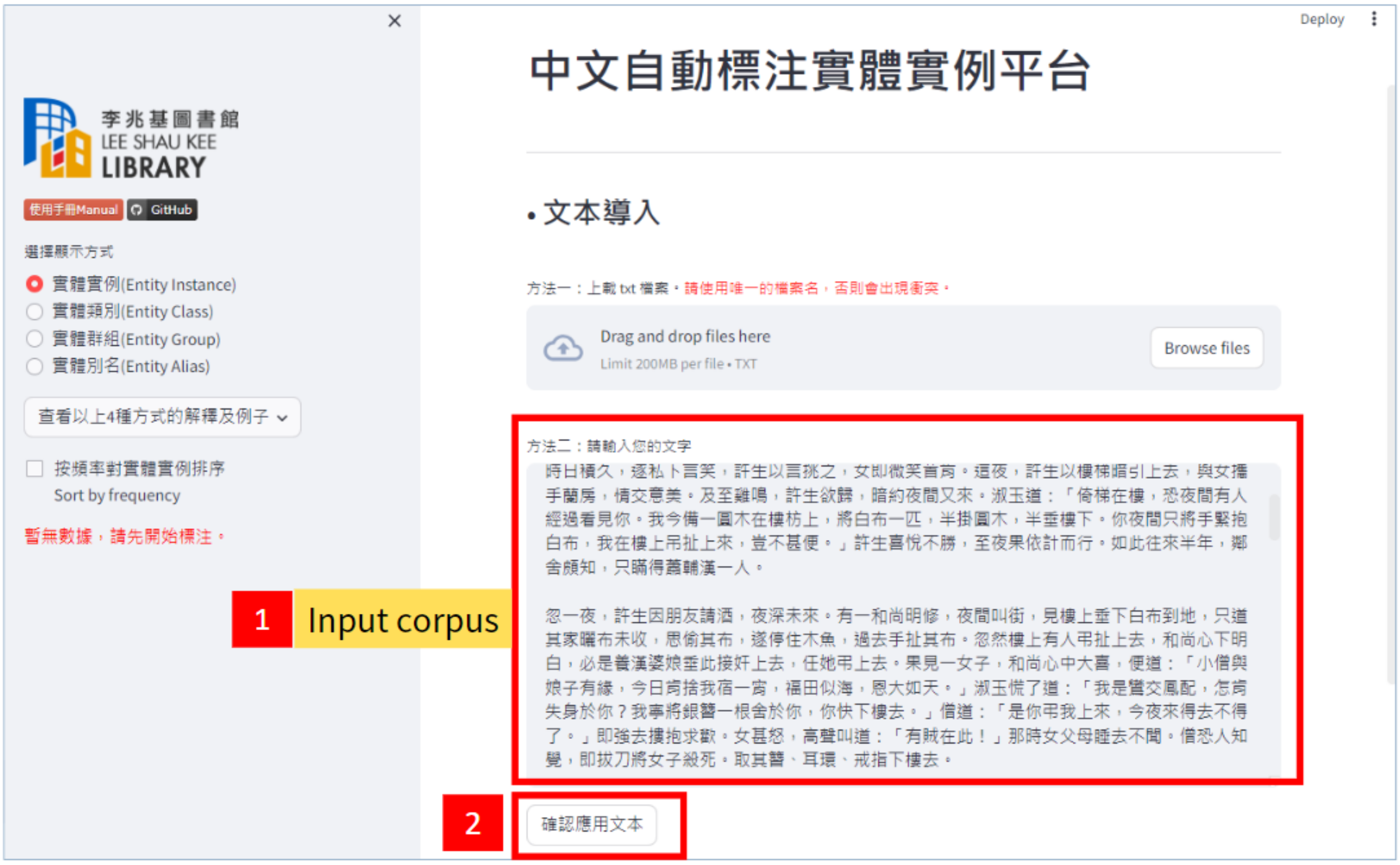

### 5.3.2 Auto Named-Entity Recognition (NER)

After clicking the “Start Auto-Annotation” button to automatically recognize entity instances (see Figure 5), some words will be highlighted, with each color representing a different entity class. The sidebar on the left shows the frequency of each entity instance. The tool provides sorting function based on frequency, helping users quickly identify the entity instances that are frequently appeared. The delete button allows users to remove any entity instances as needed (see Figure 6).

## Figure 5

*Auto Annotation*

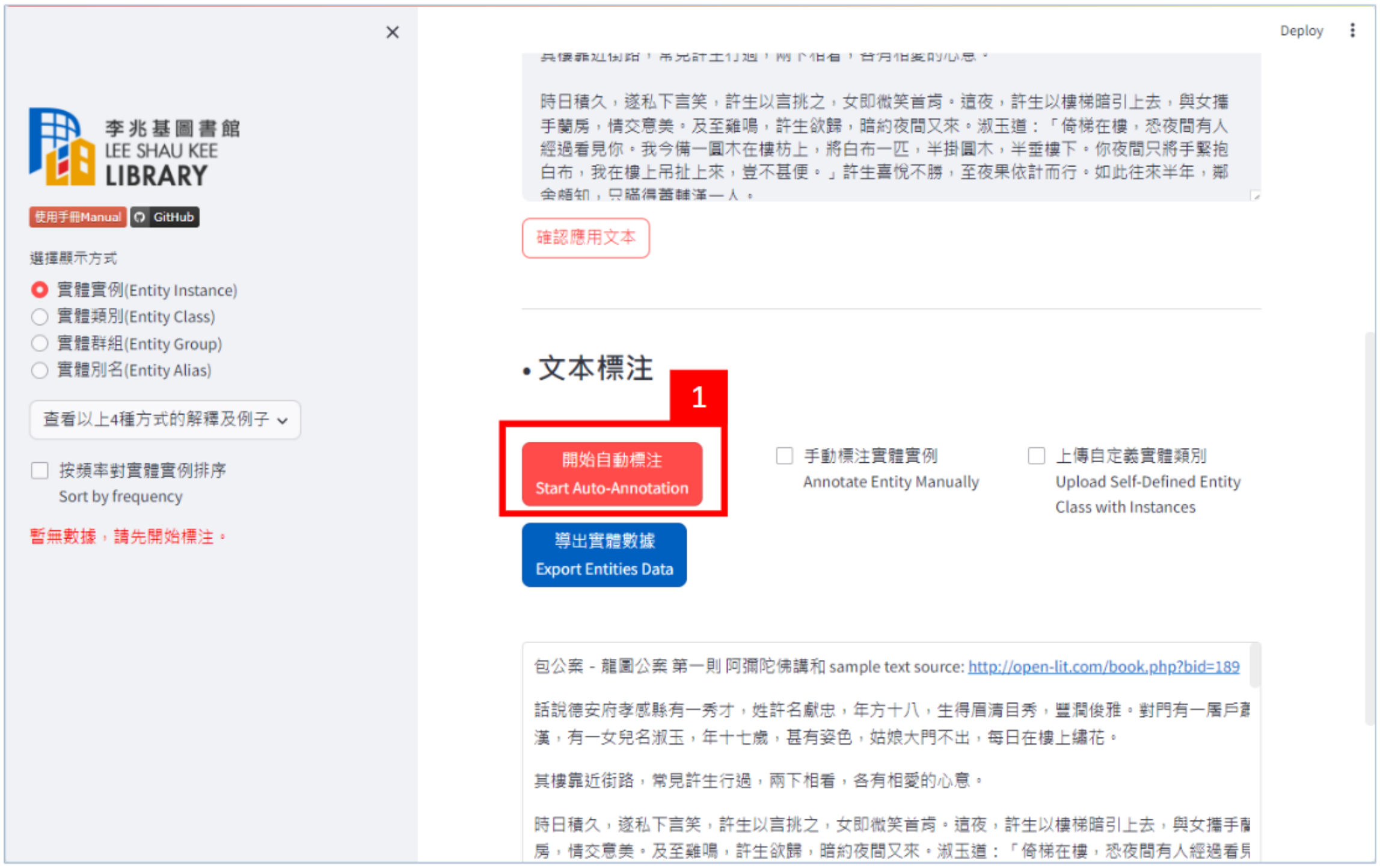

## Figure 6

*Entity Instances Display*

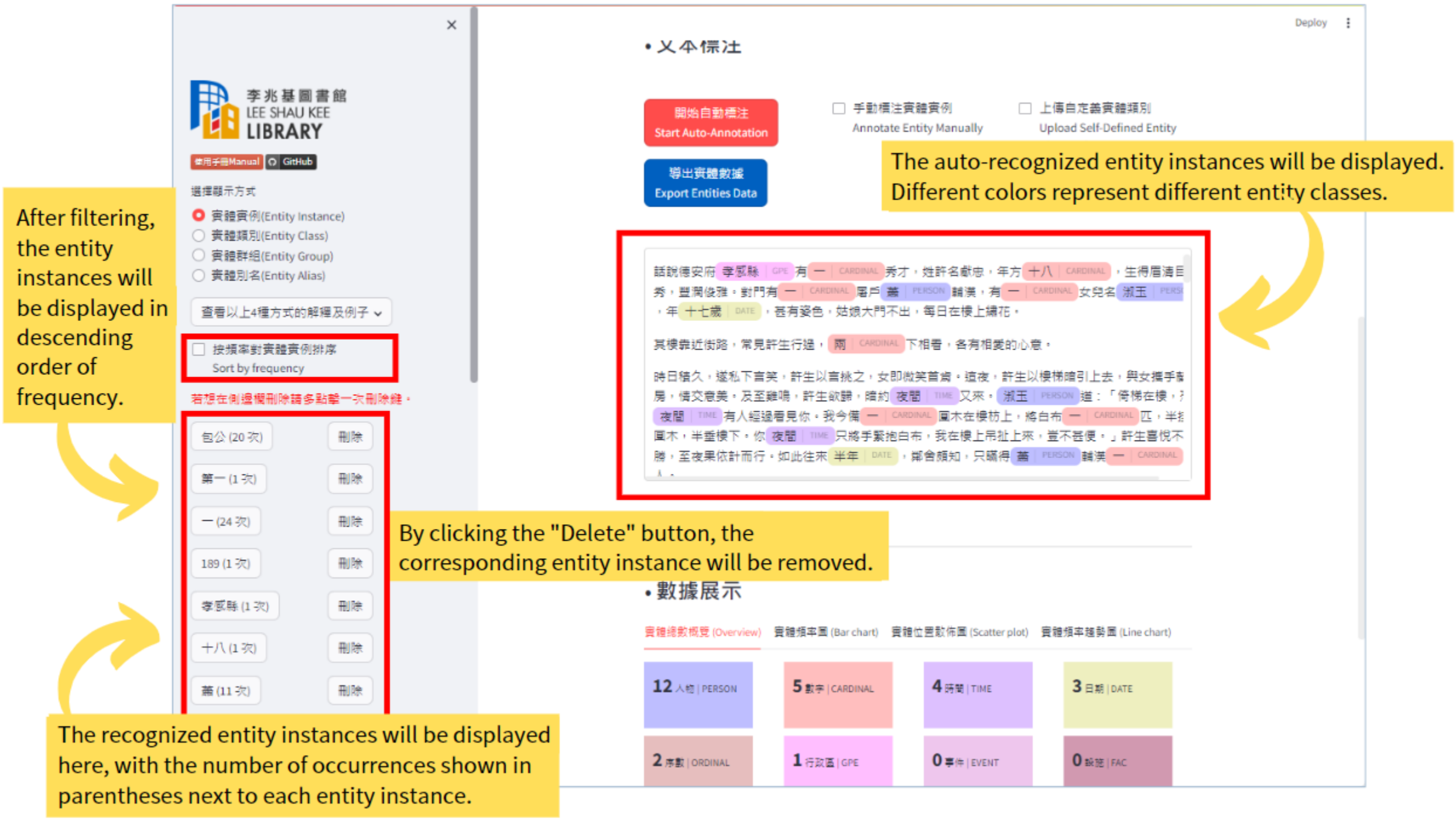

### 5.3.3. Data Management

a. Manually Add and Delete Entity Instance

Users are allowed to mark and delete entity instances in case the model omits or makes wrong decisions on some specific terms. See Figure 7 and 8 for a concrete example. It is obvious that "獻忠" belongs to the "PERSON" entity class but was not recognized by the model, so we need to manually annotate it. Similarly, for incorrectly recognized entity instances in the text, users can simply delete them manually. The detailed steps are shown in Figure 7, 8 and 9.

**Figure 7**

*Manually Add Entity Instance*

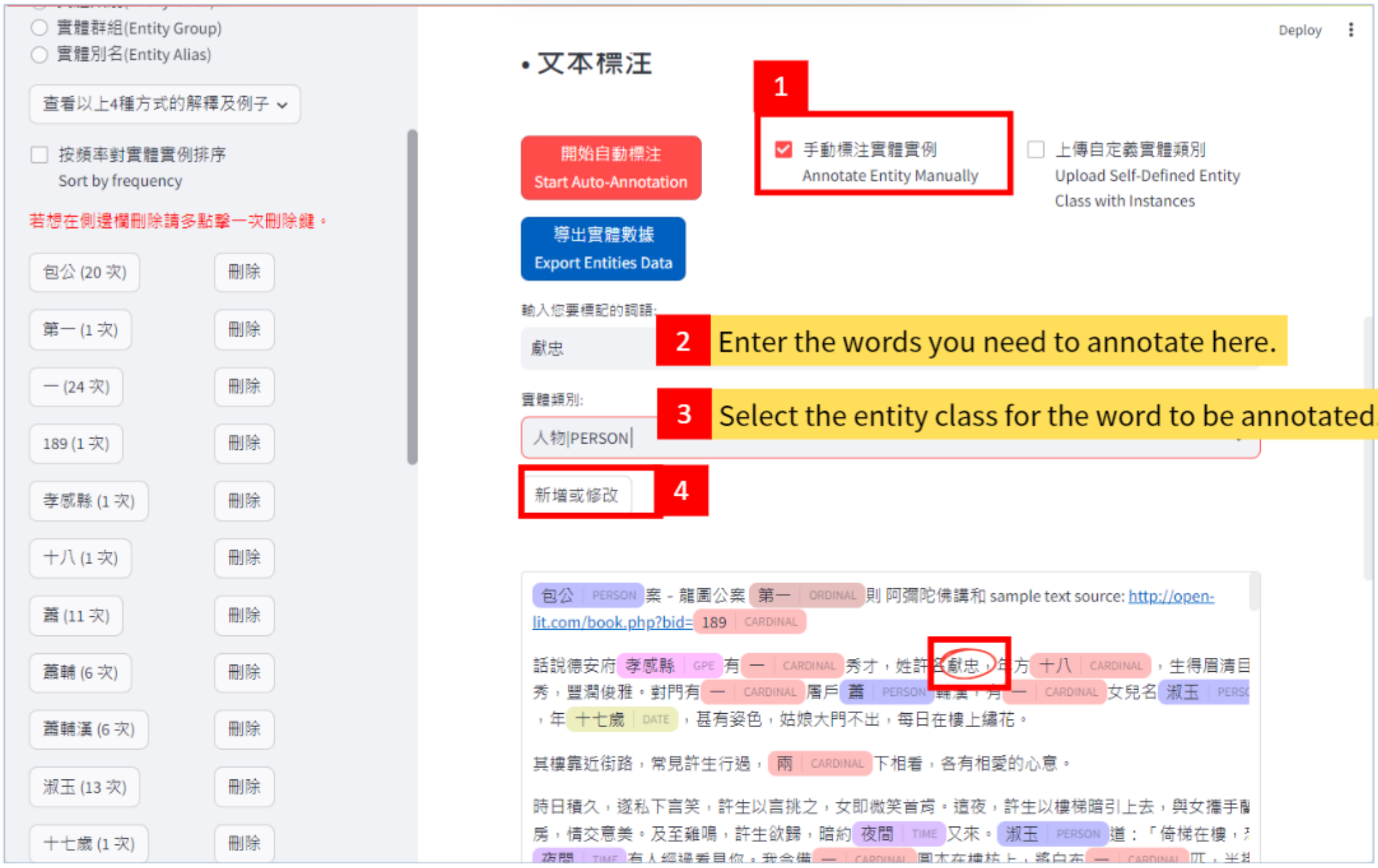

**Figure 8**

*Manually Add Entity Instance II*

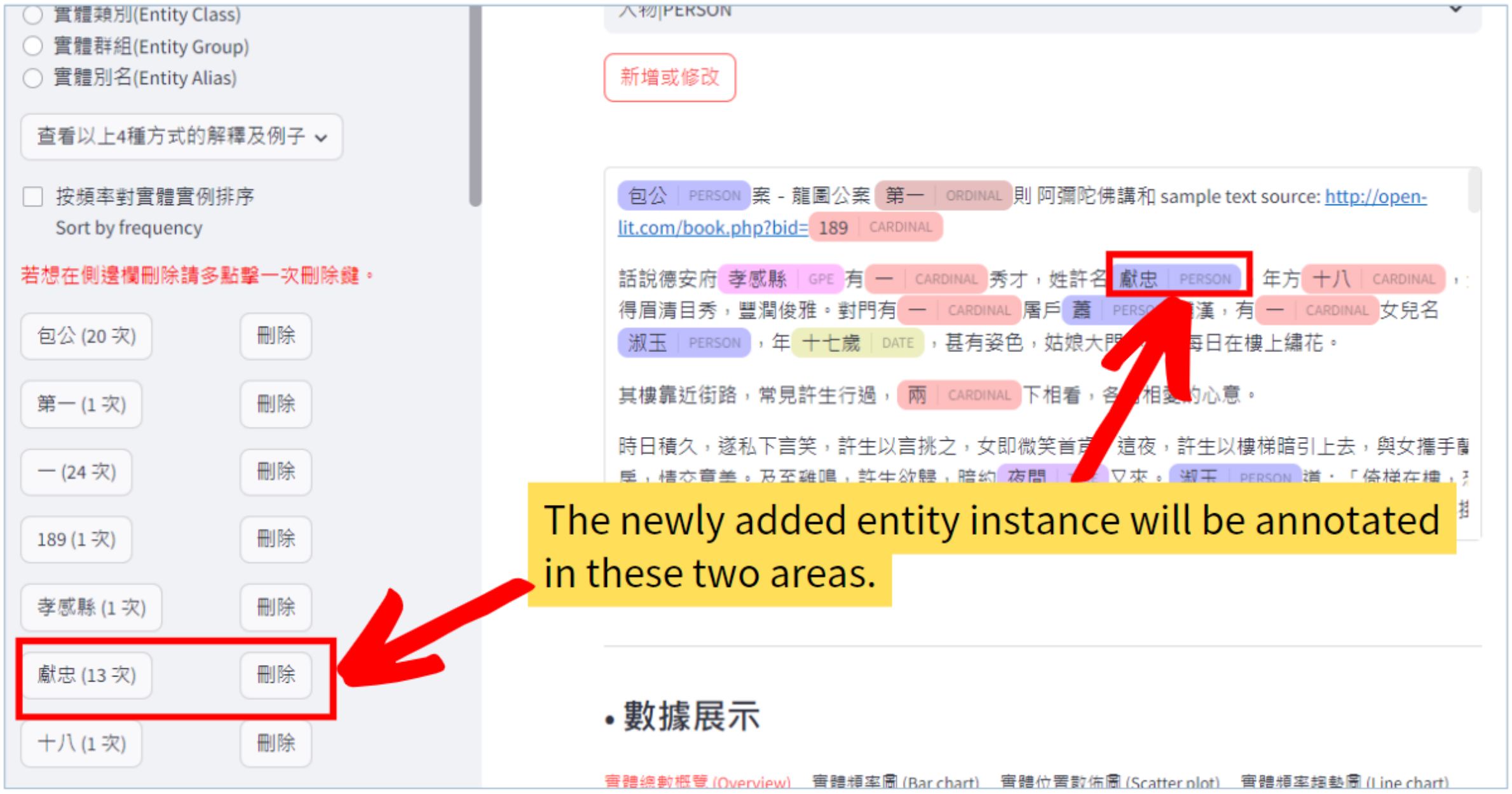

**Figure 9**

*Manually Delete Entity Instance*

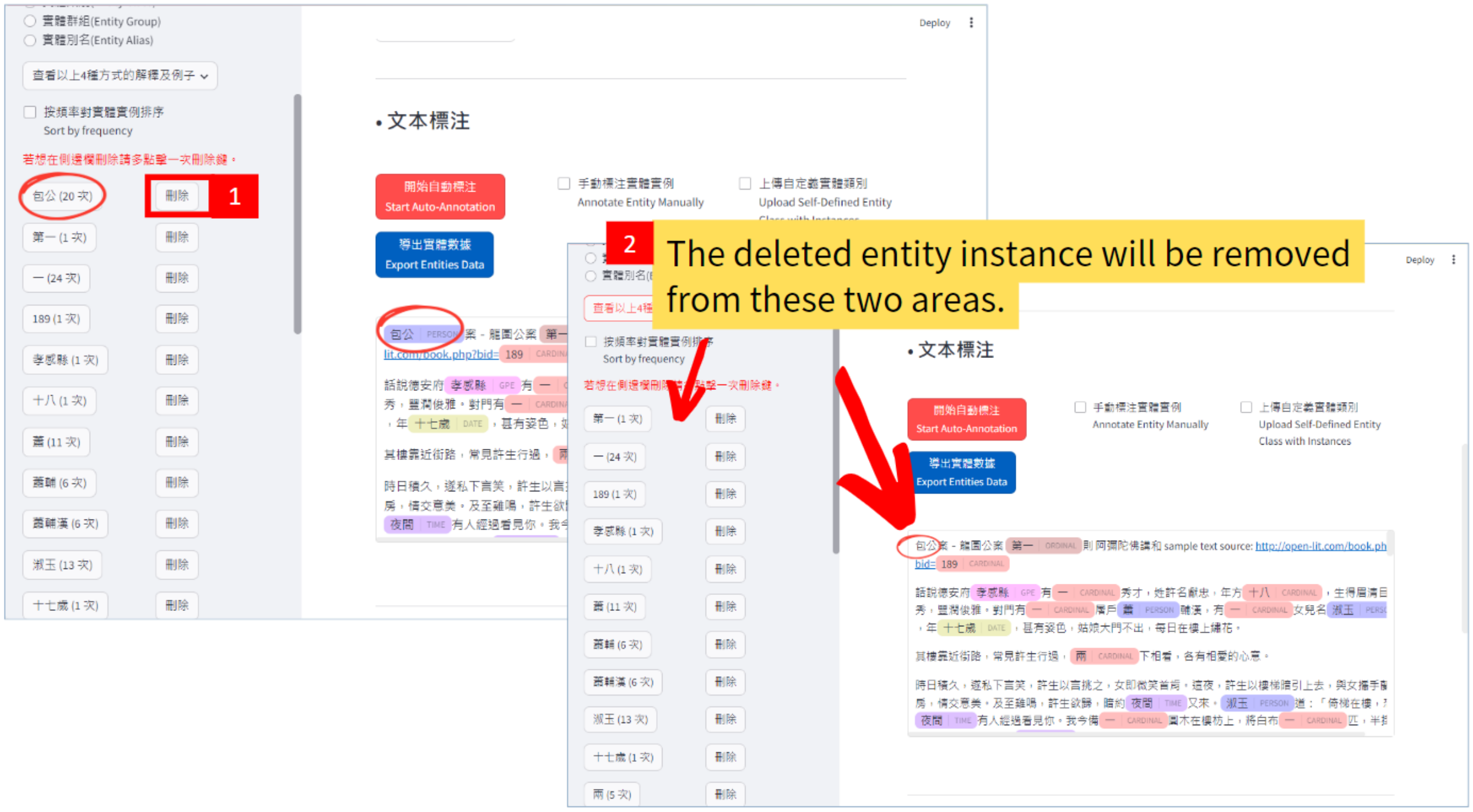

b. Manually Add and Delete Entity Class

Each entity instance can be defined as a specific entity class. In addition to providing the 18 entity classes from the CKIP model (as shown in the Table 7 below), our platform also offers the option to create new entity classes or self-defined entity classes (see Figure 10, 11 and 12). In the sidebar, we can select an entity class (such as PERSON), which allows us to see all instances of that entity class on the right-hand side of the page. This helps to observe the frequency and distribution of the characters in the article.

**Table 7**

*Default Entity Class of CKIP Model (Academia Sinica, 2019)*

| **Entity Class** | **實體類別** | **Description** |
|---|---|---|
| CARDINAL | 數字 | Numerals that do not fall under another type |
| DATE | 日期 | Absolute or relative dates or periods |
| EVENT | 事件 | Named hurricanes, battles, wars, sports events, etc. |
| FAC | 設施 | Buildings, airports, highways, bridges, etc. |
| GPE | 行政區 | Countries, cities, states |
| LANGUAGE | 語言 | Any named language |
| LAW | 法律 | Named documents made into laws |
| LOC | 地理區 | Non-GPE locations, mountain ranges, bodies of water |
| MONEY | 金錢 | Monetary values, including unit |
| NORP | 民族、宗教、政治團體 | Nationalities or religious or political groups |
| ORDINAL | 序數 | “first”, “second” |
| ORG | 組織 | Companies, agencies, institutions, etc. |
| PERCENT | 百分比率 | Percentage (including “%”) |
| PERSON | 人物 | People, including fictional |
| PRODUCT | 產品 | Vehicles, weapons, foods, etc. (Not services) |
| QUANTITY | 數量 | Measurements, as of weight or distance |
| TIME | 時間 | Times smaller than a day |
| WORK_OF_ART | 作品 | Titles of books, songs, etc. |

**Figure 10**

*Filter by Entity Class*

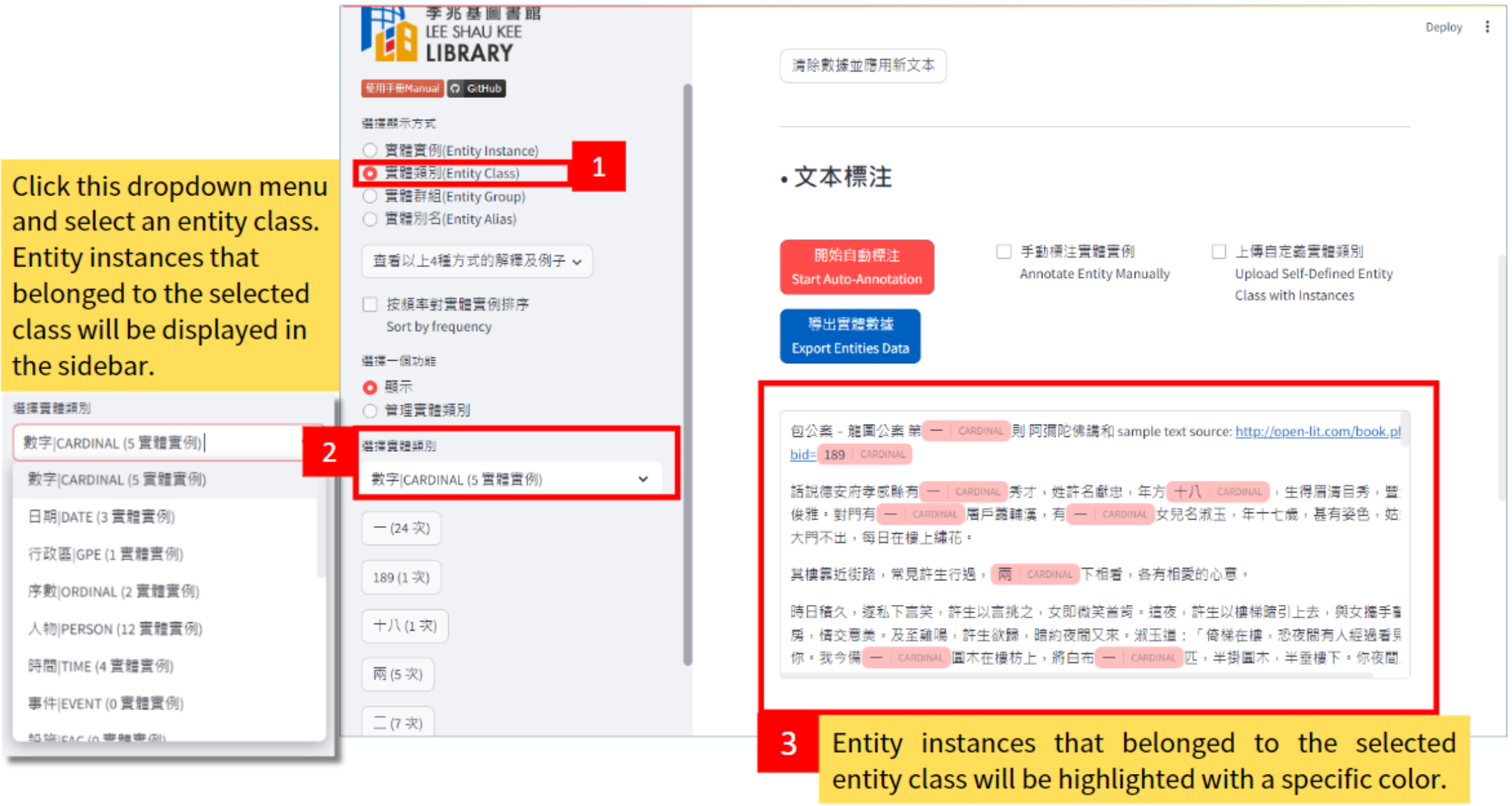

**Figure 11**

*Create New Entity Class Manually*

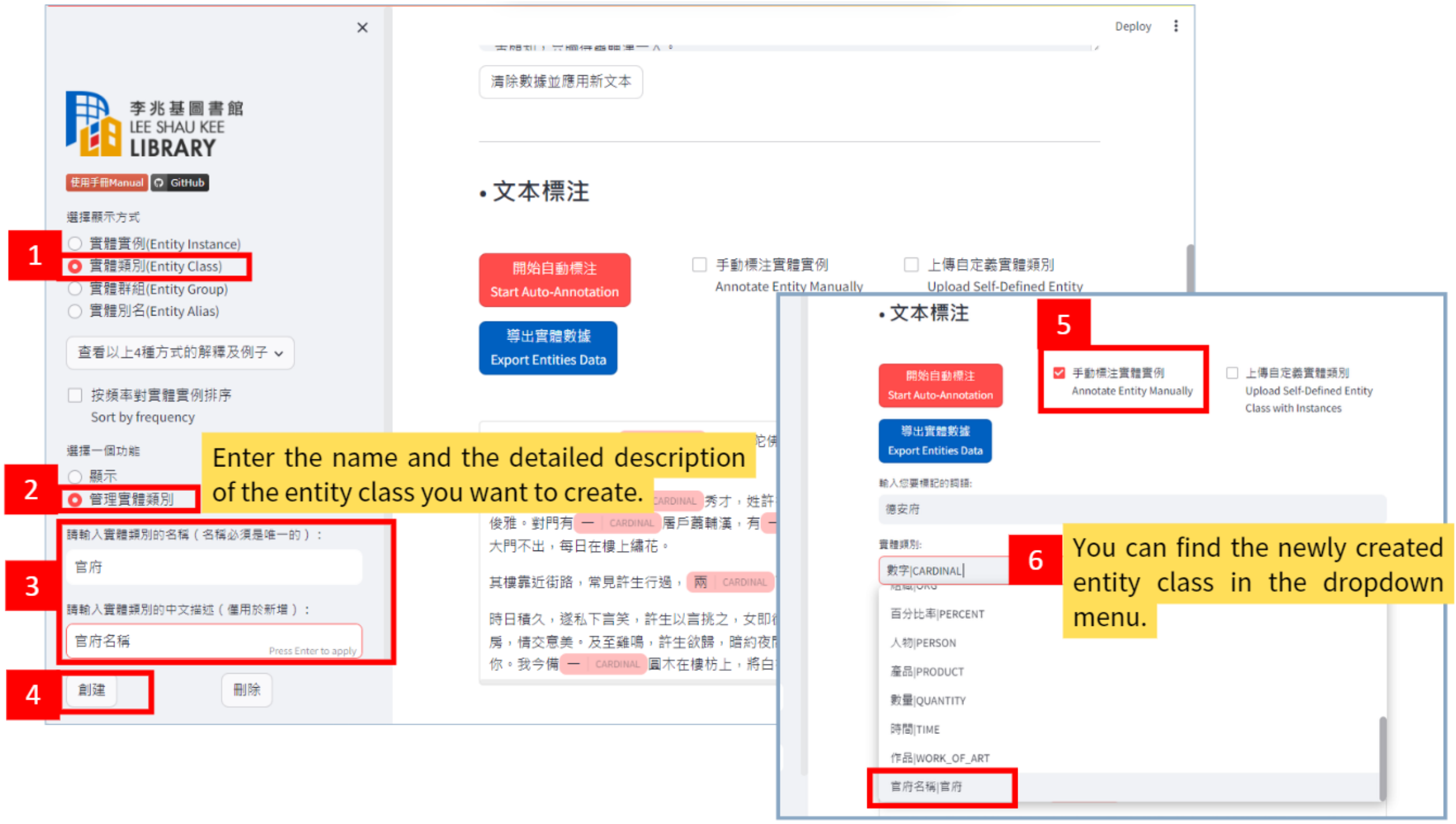

c. Manually Add and Delete Entity Group

In addition to tagging individual entities, our platform also supports grouping entity instances. Users can categorize entities into different groups based on their relevance and shared characteristics. This feature helps users better understand the relationships between entities and allows for higher-level analysis and research of these entities. For example, if we want to specifically study the main characters of the case of “Bao Gong Case”, we can create a new entity group that includes the two characters “許獻忠 (Xu Xianzhong)” and “蕭淑玉 (Xiao Shuyu)”. Then, by filtering to select this group, we can focus only on the frequency and distribution of “許獻忠 (Xu Xianzhong)” and “蕭淑玉 (Xiao Shuyu)” as one group for analysis (see Figure 12).

**Figure 12**

*Create New Entity Group*

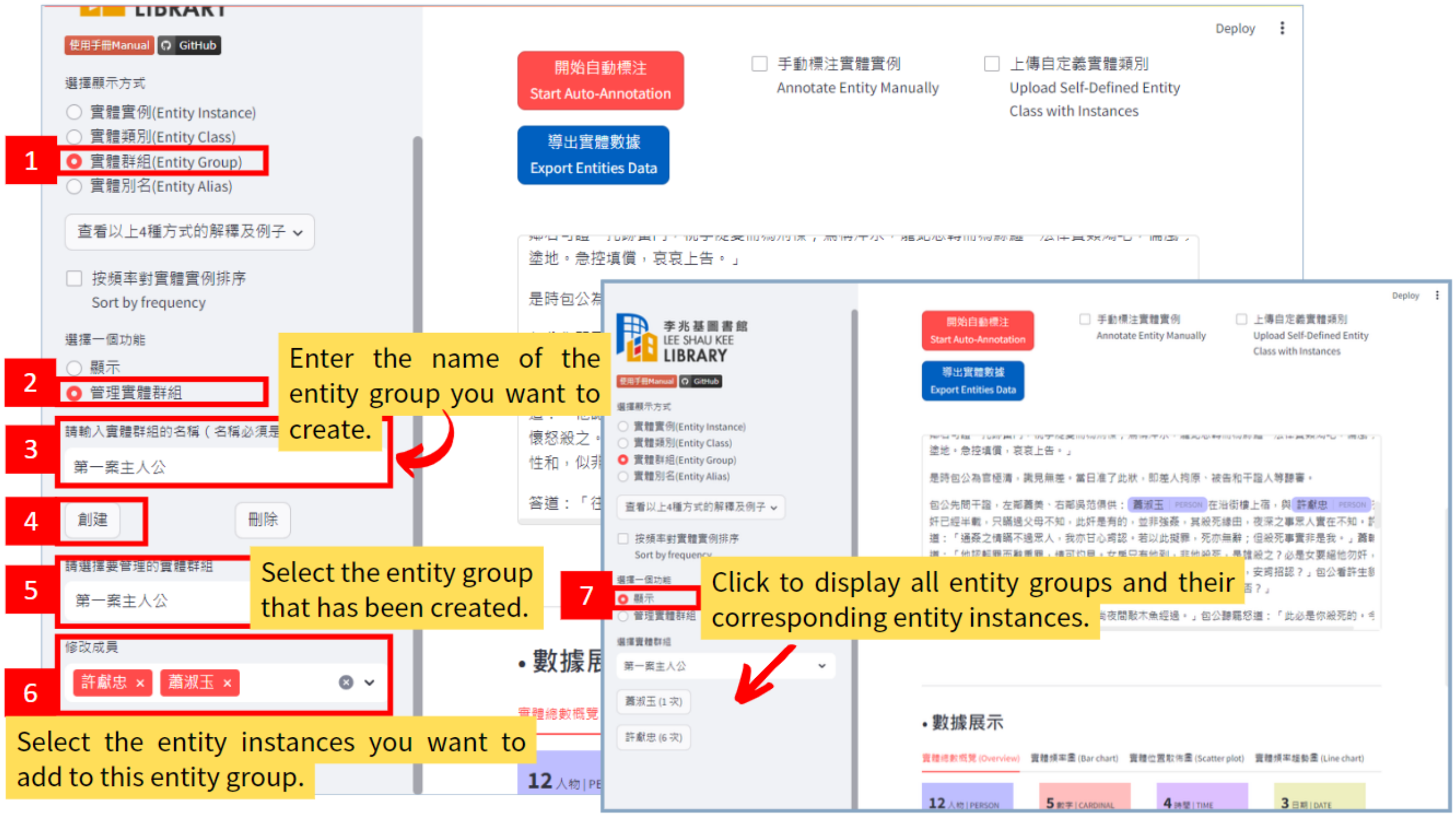

d. Manually Add and Delete Entity Alias

Our platform also supports merging entity instances that refer to the same object into a single alias and viewing their frequency accordingly. This helps to clearly present duplicate concepts in the data, allowing for better analysis of the relationships and distribution of entities. In this example text, we can find that some entity instances have the same meaning. For example, “許獻忠 (Xu Xianzhong)”, “獻忠 (Xianzhong)” and “許生 (Xu Sheng)” refer to the same person. Therefore, we can merge these three entity instances into one alias (see Figure 13). In this case, we can combine the frequency and distribution of these three terms for subsequent analysis.

**Figure 13**

*Create Entity Alias*

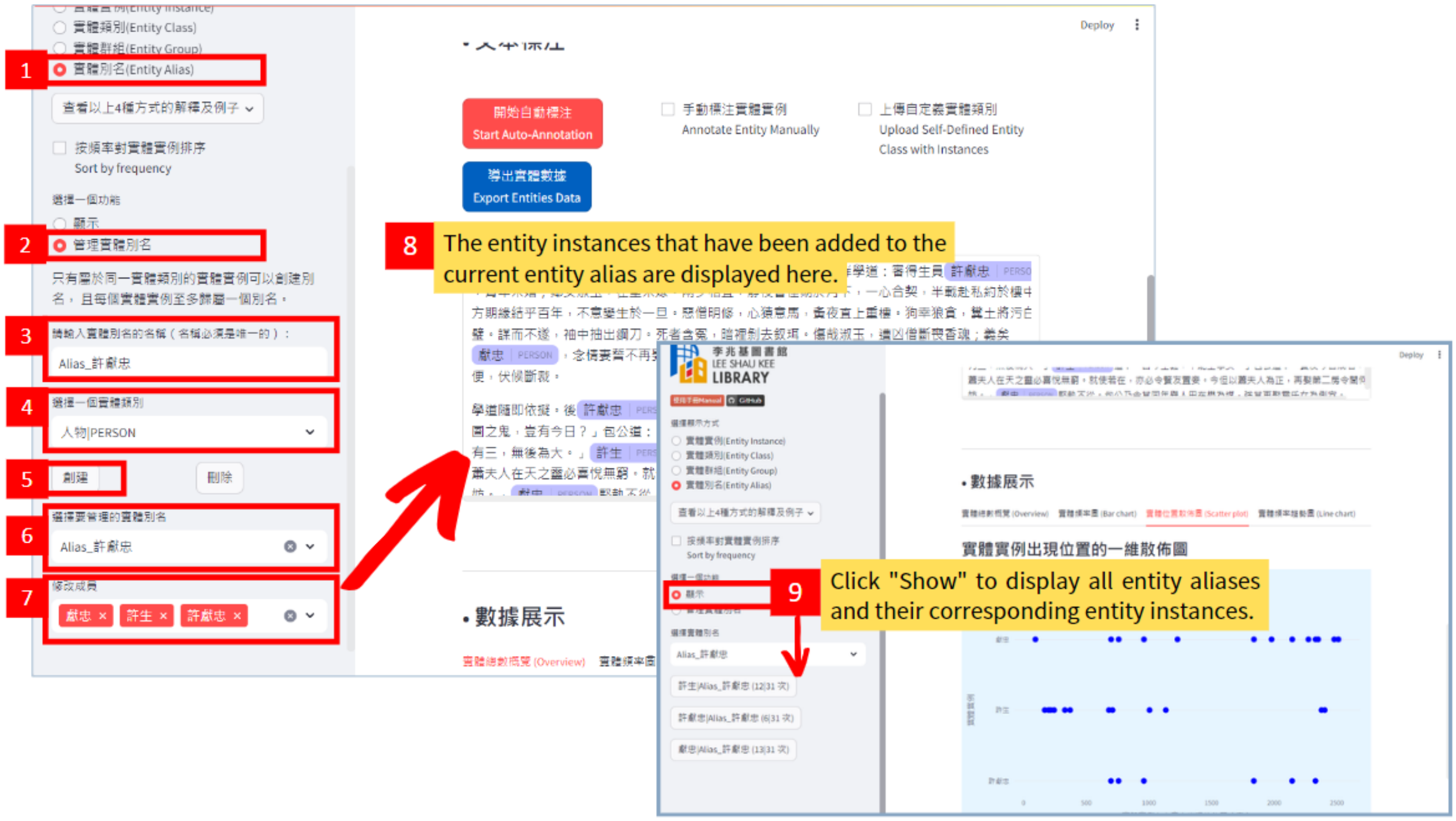

### 5.3.4. Data Visualization

To provide a more intuitive display of entity occurrence frequency and distribution, our platform offers data visualization features. Users can view charts located at the bottom of the right-hand side of the main screen to observe the frequency and position of entities within a text. This aids users in gaining a better understanding of how words are appeared throughout the document.

a. Overview

The first graph is an overview of the amount of entity instances in every entity class. Different entity classes are represented by colored blocks, with each color corresponding to a specific entity class. The numbers within these blocks indicate the number of entity instances associated with each entity class. For example, in Figure 14, there are 13 entity instances belong to the PERSON entity class, and 5 entity instances belong to the CARDINAL entity class, and so on.

**Figure 14**

*Entity Count Overview*

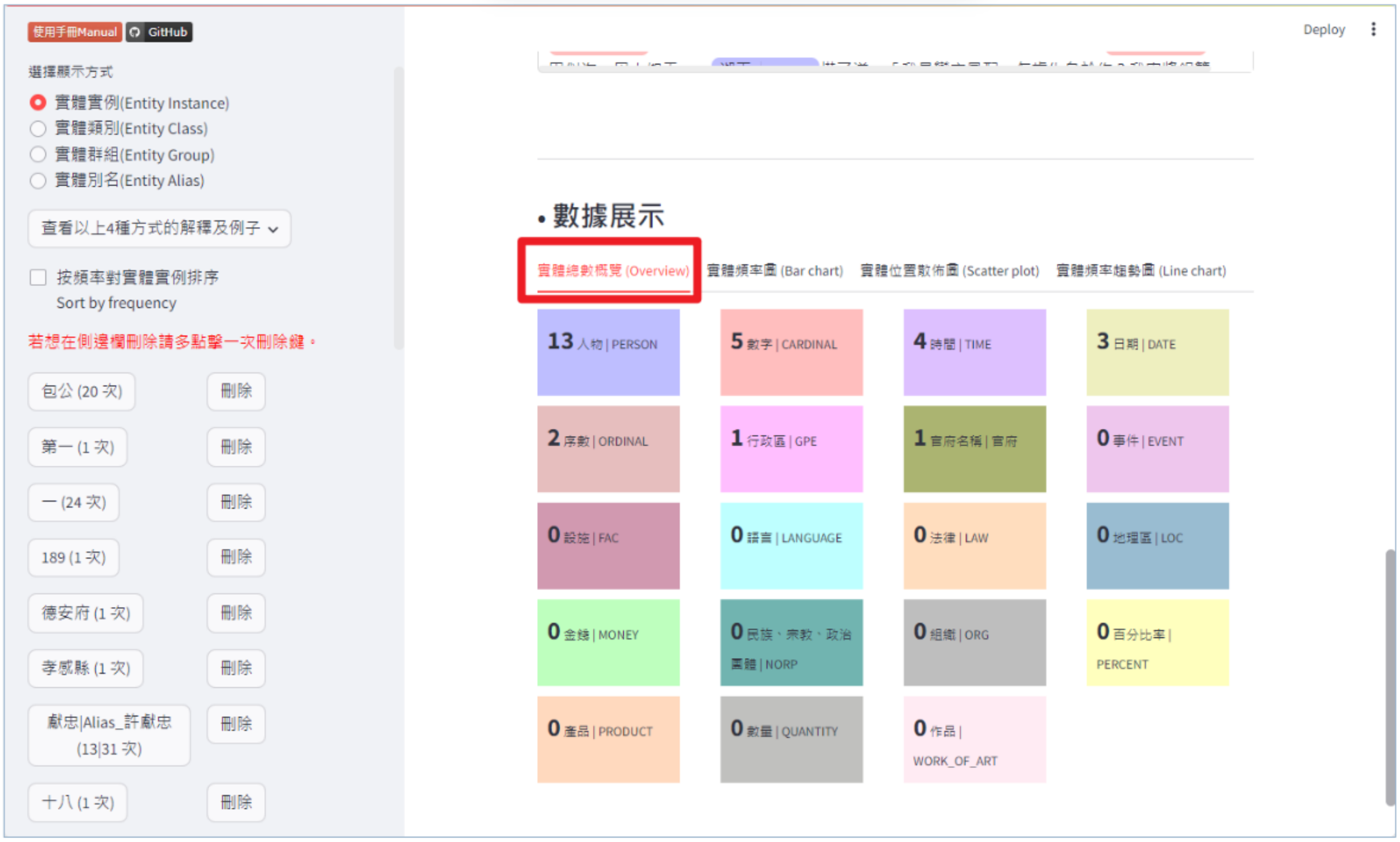

b. Bar Chart

The second tab features a bar chart. It shows the frequency of entity instances in descending order, as shown in Figure 15. Users can use filters to select specific entity class or group on the sidebar to only focus on the entity instance in the corresponding set. For example, if we select the PERSON entity class, the chart will only display the entity instances that belong to PERSON. Entity aliases can be combined or separated based on the user's requirements. In this way, "許獻忠 (Xu Xianzhong)", "獻忠 (Xianzhong)" and "許生 (Xu Sheng)" will not appear repeatedly.

**Figure 15**

*Bar Chart for Frequency*

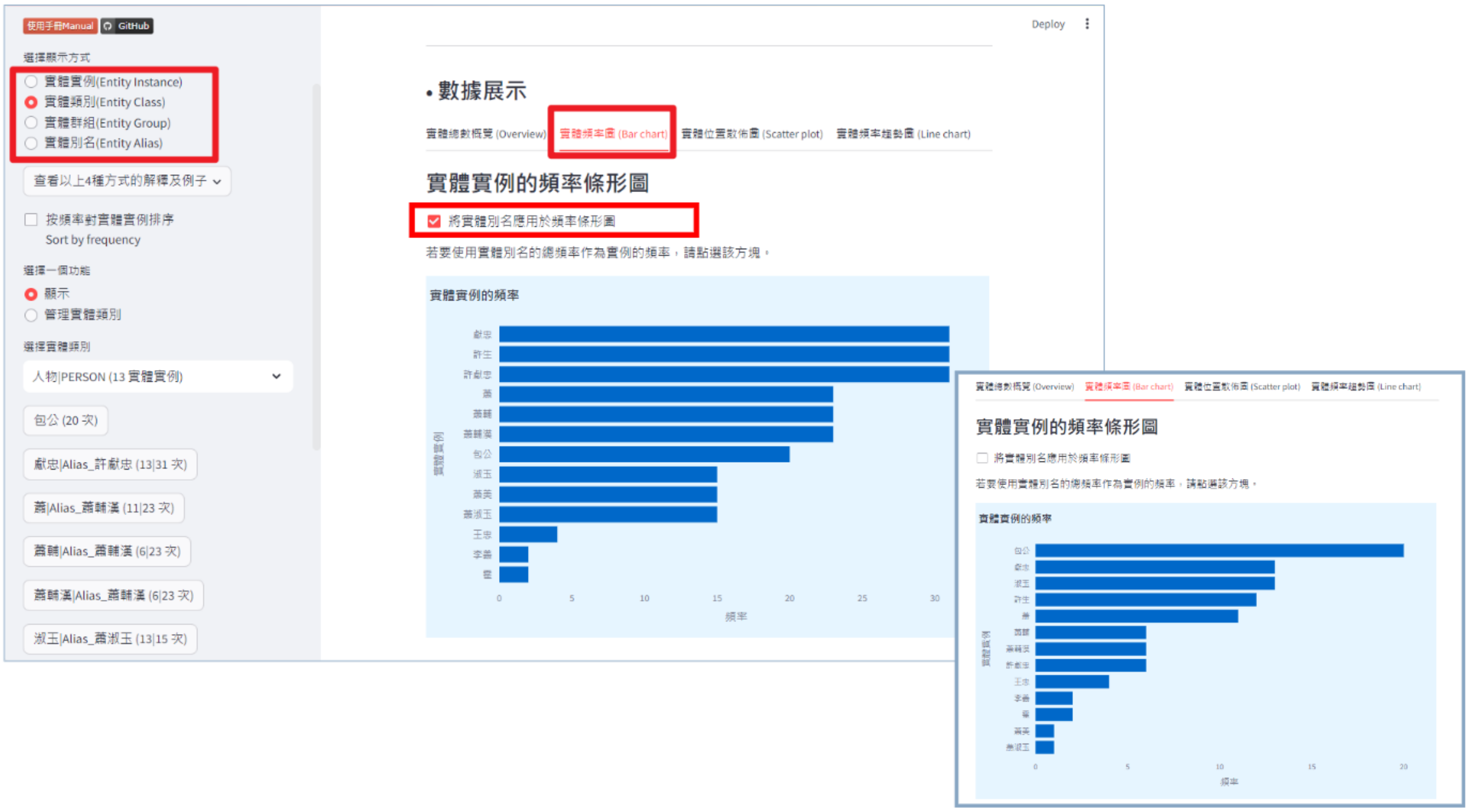

c. Scatter plot

The third tab features a scatter plot. This chart displays the position of each entity instance as it appears in the text. Within this approach, users can intuitively understand the distribution of the target instance and get an insight into the main content of each part of the article. For example, in Figure 16, we can see that “許獻忠 (Xu Xianzhong)” primarily appears in the first half of the article as a key figure, while "包公 (Bao Gong)" as the judge, mainly appears in the second half of the article.

**Figure 16**

*Scatter Plot for Position of Entity Instances*

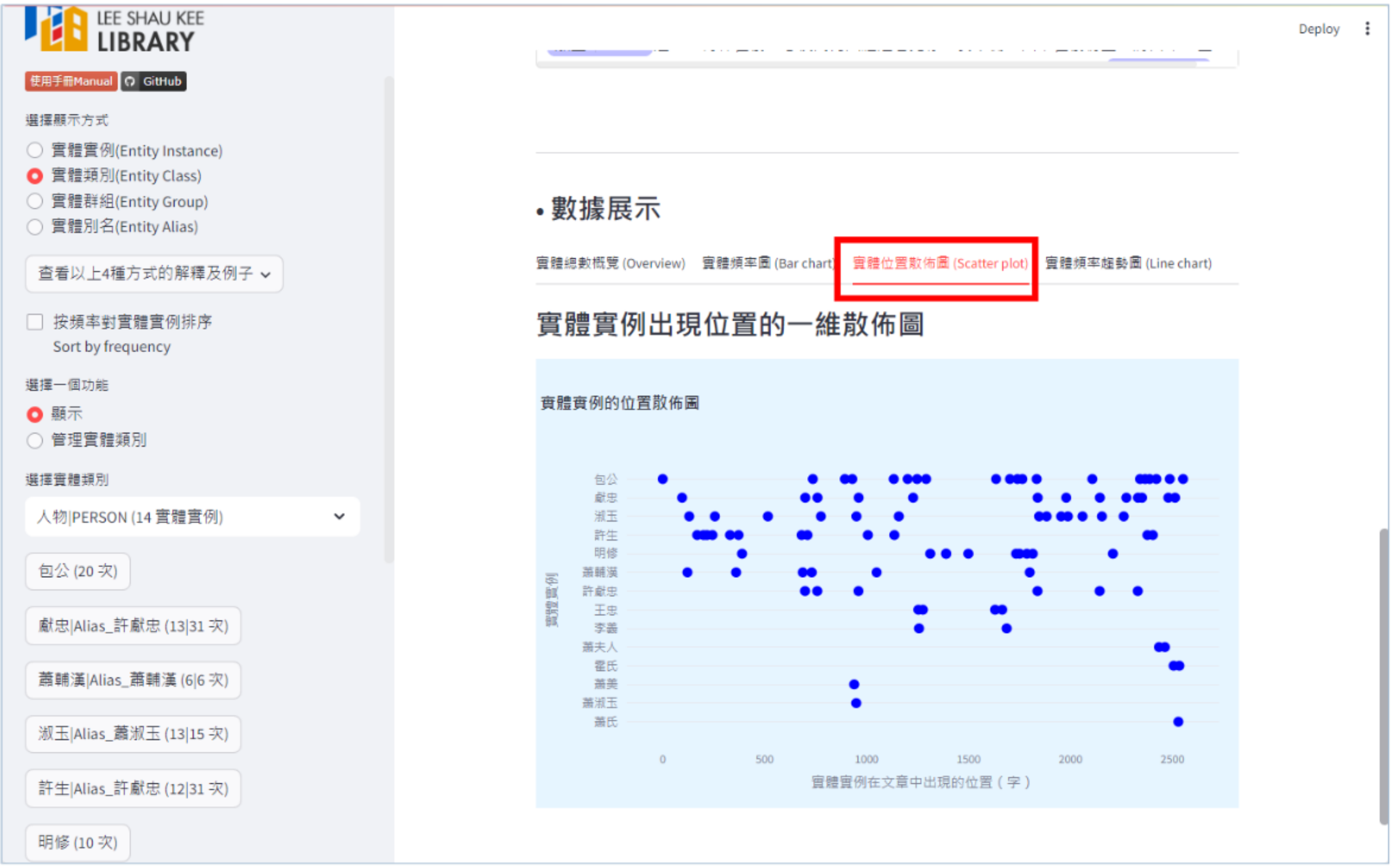

#### 5.3.5 Data Export

Having the ability to export data is vital for subsequent analysis, as it provides flexibility for users to utilize the information in multiple ways. Our tool supports entity data export functionality. Users can export the annotated entity data into CSV formats (see Figure 17). This capability allows for additional processing and analysis with other software or platforms. The data.zip file we offered for export contains three CSV files, with each row in these files containing a unique ID (see Table 7).

**Table 7**

*Export Data CSV Format*

| **Entity.csv** | **Alias.csv** | **Group.csv** |
|---|---|---|
| • ID_E<br>• Entity<br>• EntityClass<br>• Frequency<br>• Relations<br>• Relation_Name | • ID_A<br>• AliasName<br>• AliasClass<br>• AliasFrequency<br>• AliasMembers<br>• AliasMembers_Name | • ID_G<br>• GroupName<br>• GroupFrequency<br>• GroupMembers<br>• GroupMembers_Name |

**Figure 17**

*Data Export*

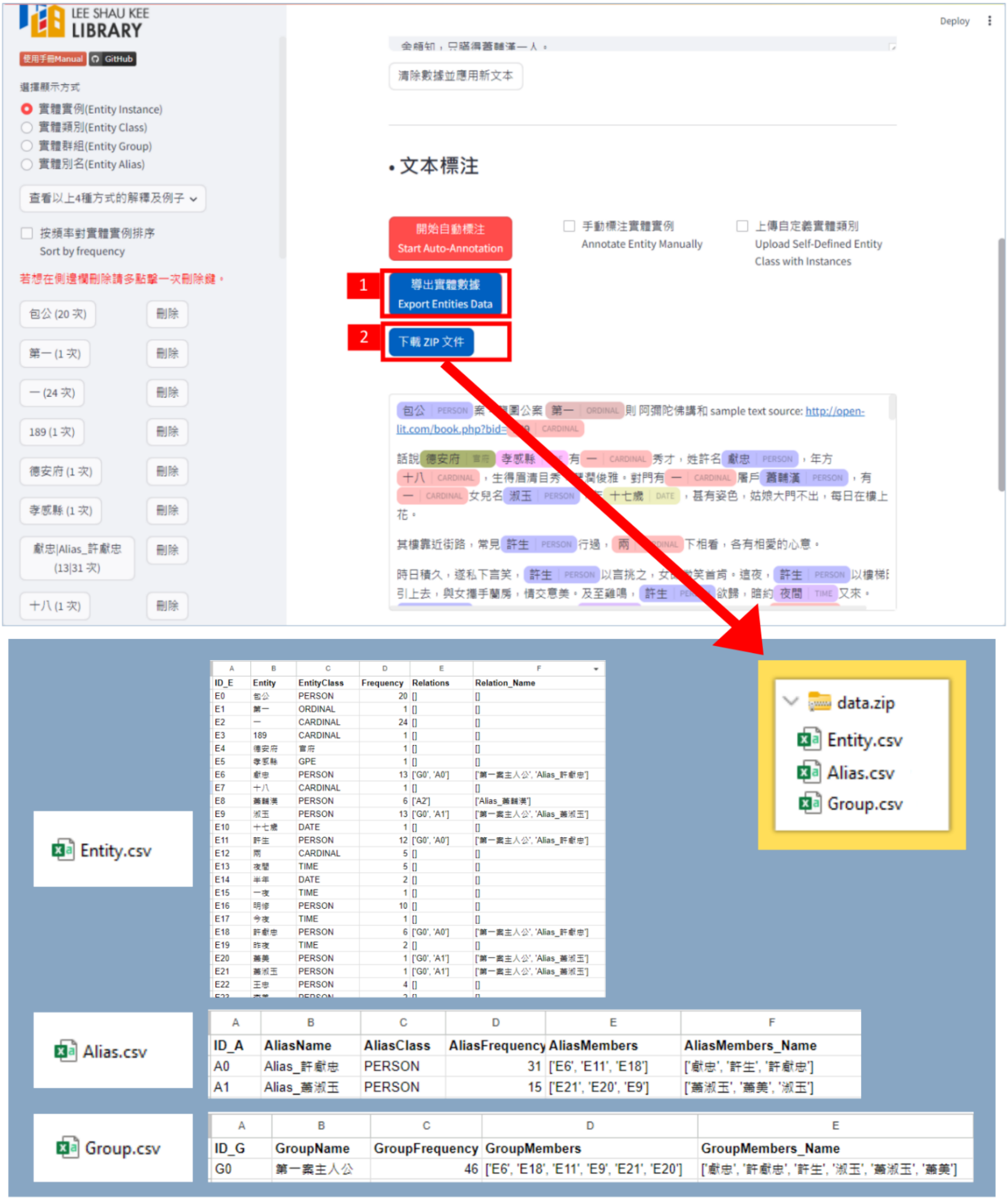

## 5.4 Application of Our Tool – Scenario Example (Multiple Files)

In the previous section, we demonstrated the use of our tool for processing and analyzing a single document. In this section, we will use chapters 59, 60 and 61 of "*西遊記 (Journey to the West)*" (which is related to the plot of *三借芭蕉扇 Borrowing the Banana Fan Three Times*, full text from: http://open-lit.com/book.php?bid=14) as example to illustrate the case for multiple files.

Most of the functions are the same for the cases for multiple files, with some of the functions can be applied to all the uploaded files while some functions can only be applied to the currently viewed file. Specifically, as stated in Table 6, functions applicable to all uploaded files are: entity instance automatic recognition, customized definition upload, management of entity instance and entity class, while those only affecting the currently viewing file are: management of entity group and entity alias, data visualization and data export. Particularly, the line chart for entity frequency across files (see Figure 26) is available only in the case of multiple files.

### 5.4.1. Corpus Import (multiple files)

File uploader must be used to import multiple files, and the names of the TXT files should be carefully designed to cooperate with the alphabetical ordering of the uploader. An additional drop-down box is used to switch currently viewed file (see Figure 18).

**Figure 18**

*Import Corpus Via Uploading Txt File(s)*

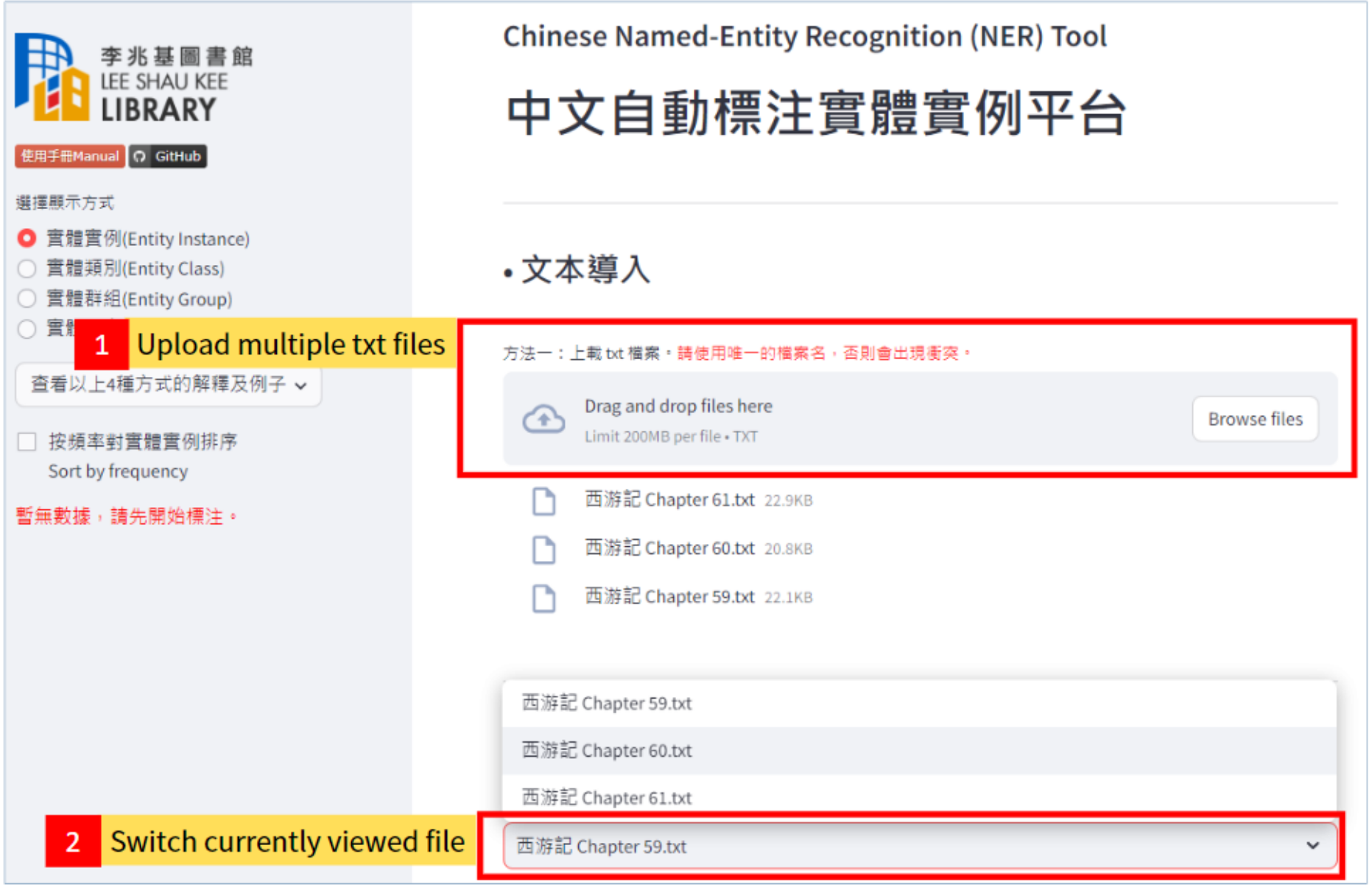

### 5.4.2 Entity Instance Automatic Recognition

This functionality will be applied to all files automatically when execute in one of the files (see Figure 19).

**Figure 19**

*Auto Annotation*

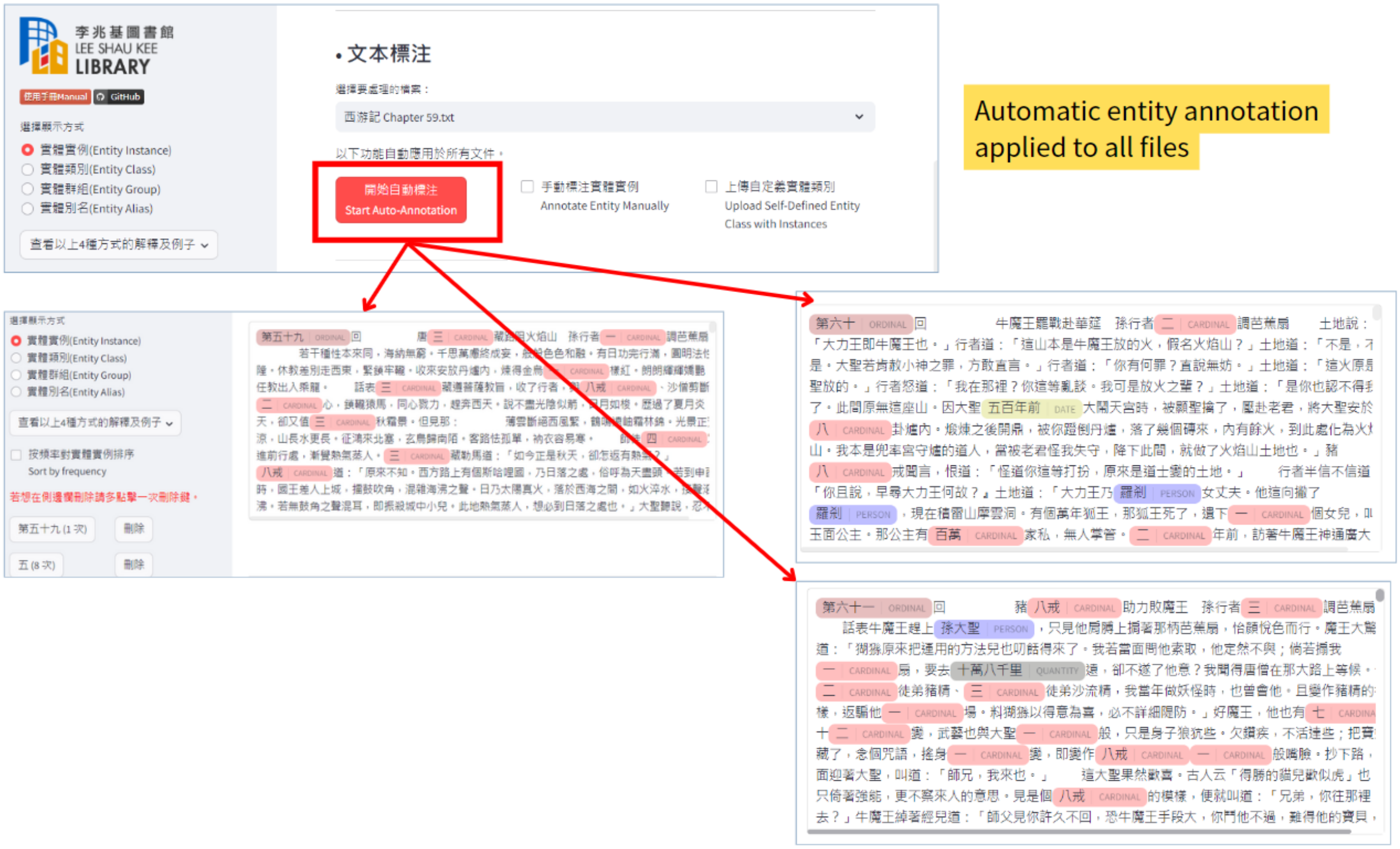

### 5.4.3 Entity Instance Management

Manual correction of entity instance is performed in the same as in the case of single file. Modification in any one of the files will be applied to all the files.

Figure 20 shows common situation to use these functions: deleting a useless entity "一" and adding a missed person name "行者".

**Figure 20**

*Example of Adding and Deleting Entity Instances*

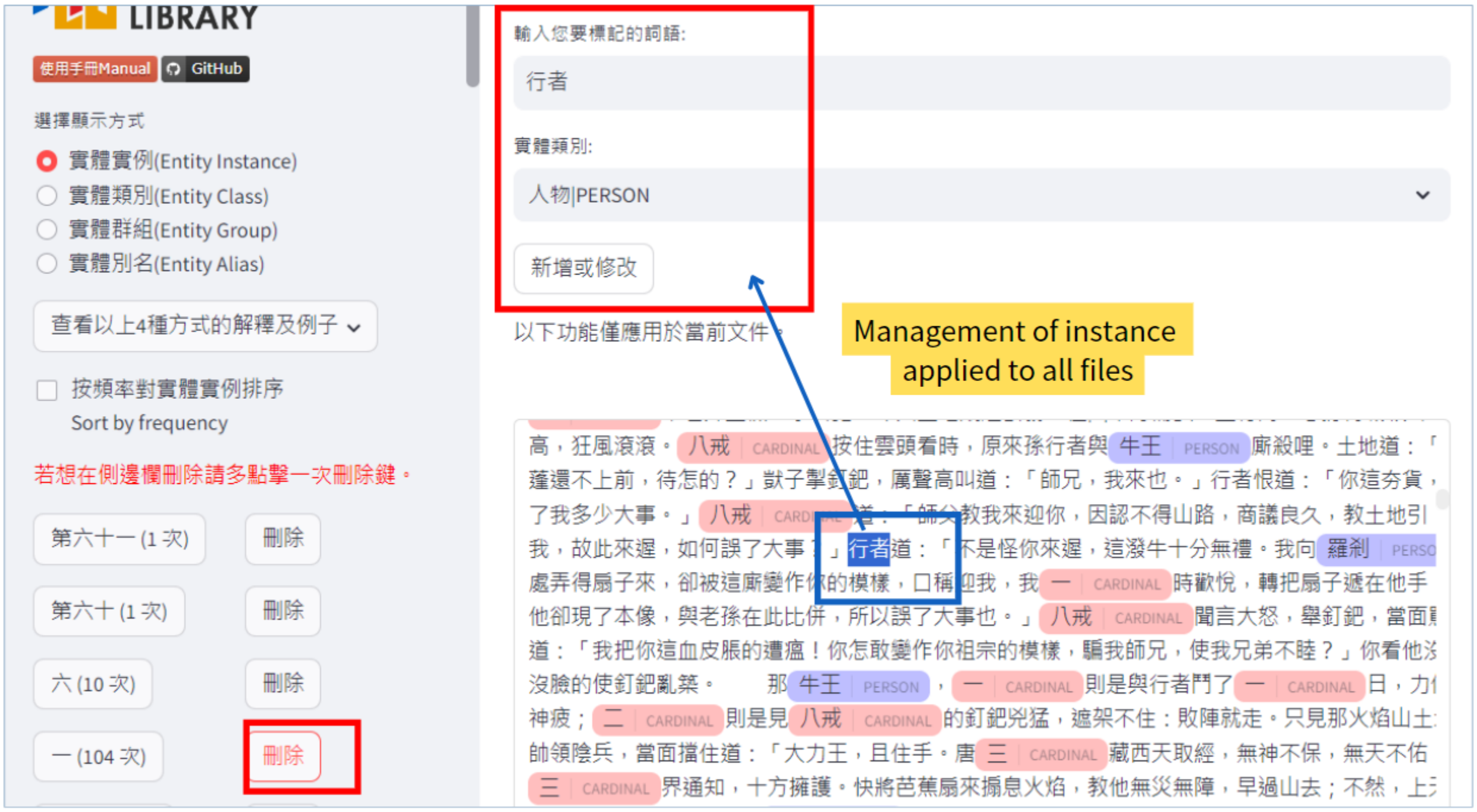

### 5.4.4 Entity Class Management

Similar procedures are performed to manage entity class and the changes will be applied to all files. For example, after going through the annotated text in our sample text, it can be noticed that there are several weapons such as “芭蕉扇” and “金箍棒” in the passages, which should belong to class “PRODUCT” according to the definition (Table 7). To be more informative, it may be better to replace “PRODUCT” with “WEAPON”. This can be achieved by using the Entity Class Deletion and Creation function (see Figure 21).

**Figure 21**

*Example of Deleting and Creating Entity Class*

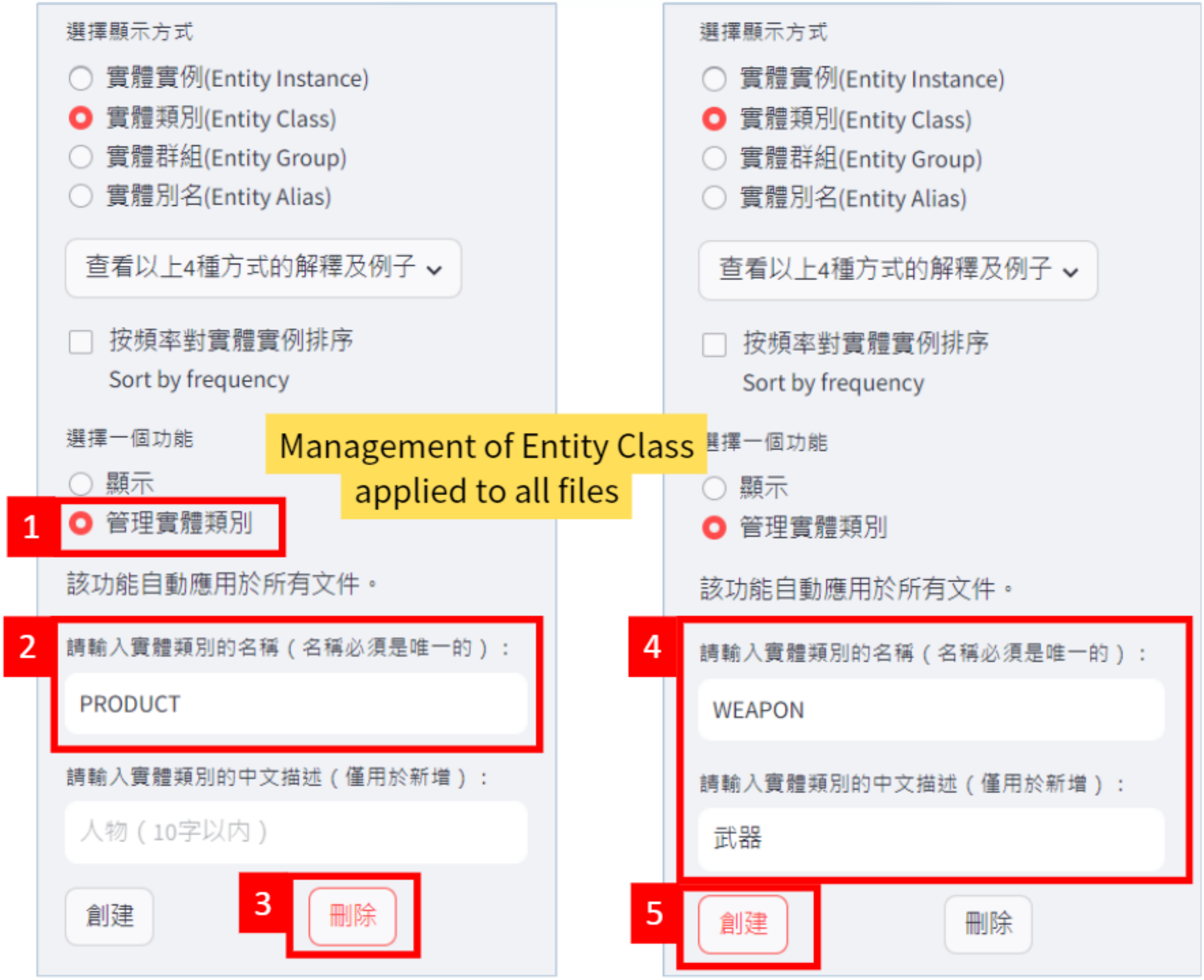

### 5.4.5 Customized Definition Upload

Similar procedures are performed to upload customized definition, and it will be applied to all files. In this example, this functionality is used to achieve a more focused analysis. It is reasonable that users might want to focus on only a few entity classes and entity instances. Several main characters, locations and weapons are chosen in this case (see Figure 22 for the details presented in the CSV). After refreshing to clean previous data and following the procedure for definition uploading, only those selected classes and entity instances will be defined.

**Figure 22**

*Example of Uploading Self-Defined Entity Class with Instances*

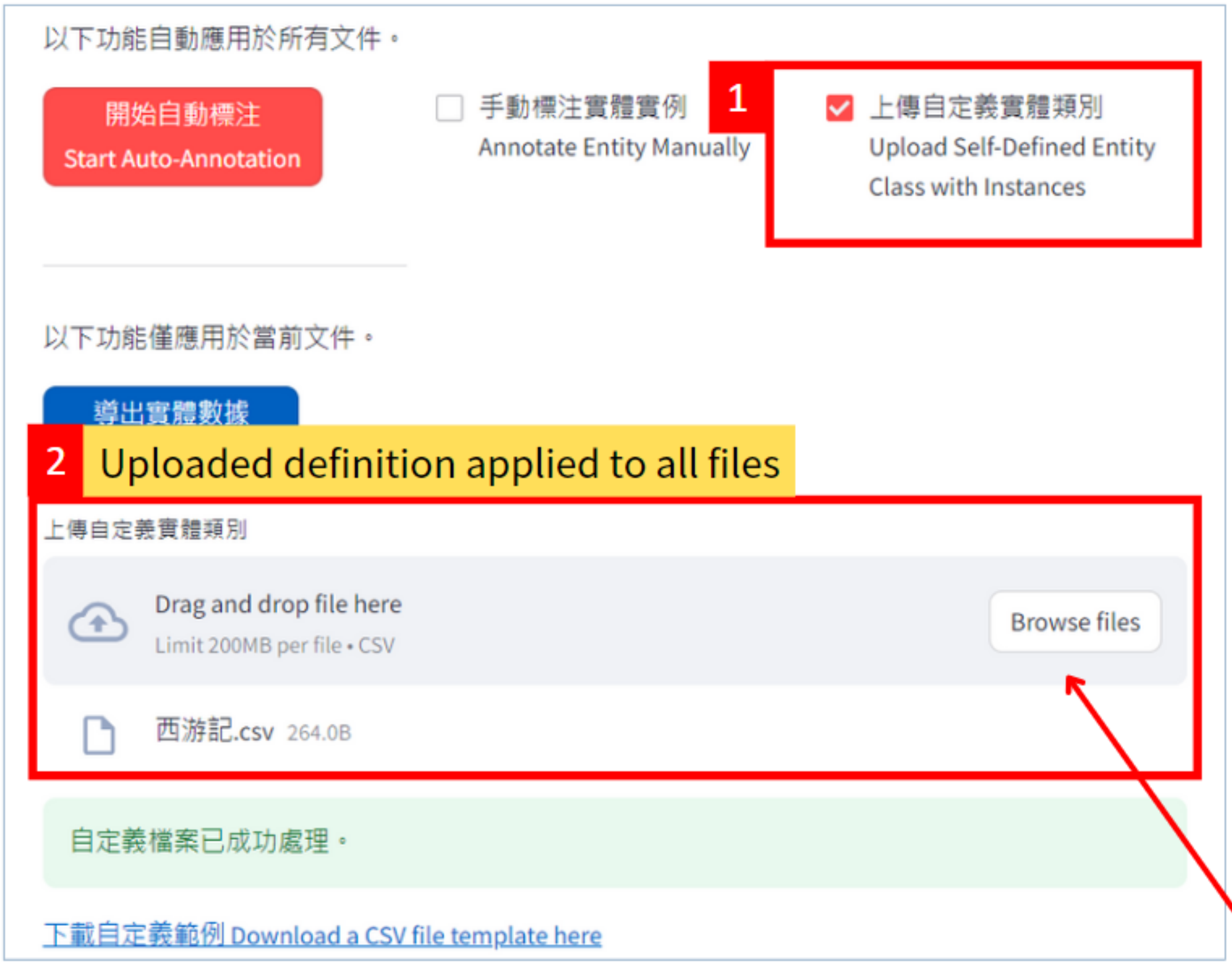

| | A | B | C |
|---|---|---|---|
| 1 | Class_Label | Class_Description | Instance_List |
| 2 | PERSON | 人物 | 三藏, 悟空, 行者, 大聖, 八戒, 沙僧, 白馬, 牛王, 牛魔王, 公主, 羅剎, 菩薩 |
| 3 | WEAPON | 武器 | 芭蕉扇, 金箍棒 |
| 4 | LOC | 地理區 | 火焰山, 翠雲山, 崑崙山, 峨眉山, 芭蕉洞 |
| 5 | | | |

### 5.4.6 Entity Alias Management and Entity Group Management

Entity Alias and entity group are defined in the exact same way as that for single file. Unlike the management of entity instance and entity class, the definition of entity alias and entity group are independent among different files. Because different files may correspond to different sets of entity instances, automatically applying to all files may introduce meaningless empty group or alias.

In this case of sample text, an entity alias named “孫悟空” can be defined to include equivalent names: “行者”, “大聖”, “悟空” and “孫悟空” (see Figure 25). Moreover, a group named “悟空和芭蕉扇” can be created to include the above entity and “芭蕉扇” to have a more focused study (see Figure 27 and 28).

### 5.4.7 Data Visualization and Display Filtering

All visualization and filtering methods mentioned in the Section 5.3.4 are available for the case of multiple files. Moreover, for multiple files, there is an additional line graph to show the variation of frequency of an entity across files. These functionalities can assist in gaining an insight into the files. Below, chapters 59-61 of the *“Journey to the West*” are used as a real case scenario to demonstrate how you can interpret the visualizations that are shown in our tool.

a. Overview

In Figure 23, Entity Class filtering is used to focus on studying "PERSON" entities (e.g. the main characters) in chapter 59. As shown in the bar chart, "行者" and "羅剎" have the highest frequencies which may indicate that they are the major characters of this chapter. As shown in the scatter plot, there are roughly four clusters of dots, which could imply the number of scenes in this chapter. Similarly in the other two chapters (see Figure 24), an overview of main characters and the interactions among them can be gained by examining the scatter plots.

**Figure 23**

*Examples of using Chapter 59 to Show How to Interpret Bar Chart and Scatter Plot in Our Tool*

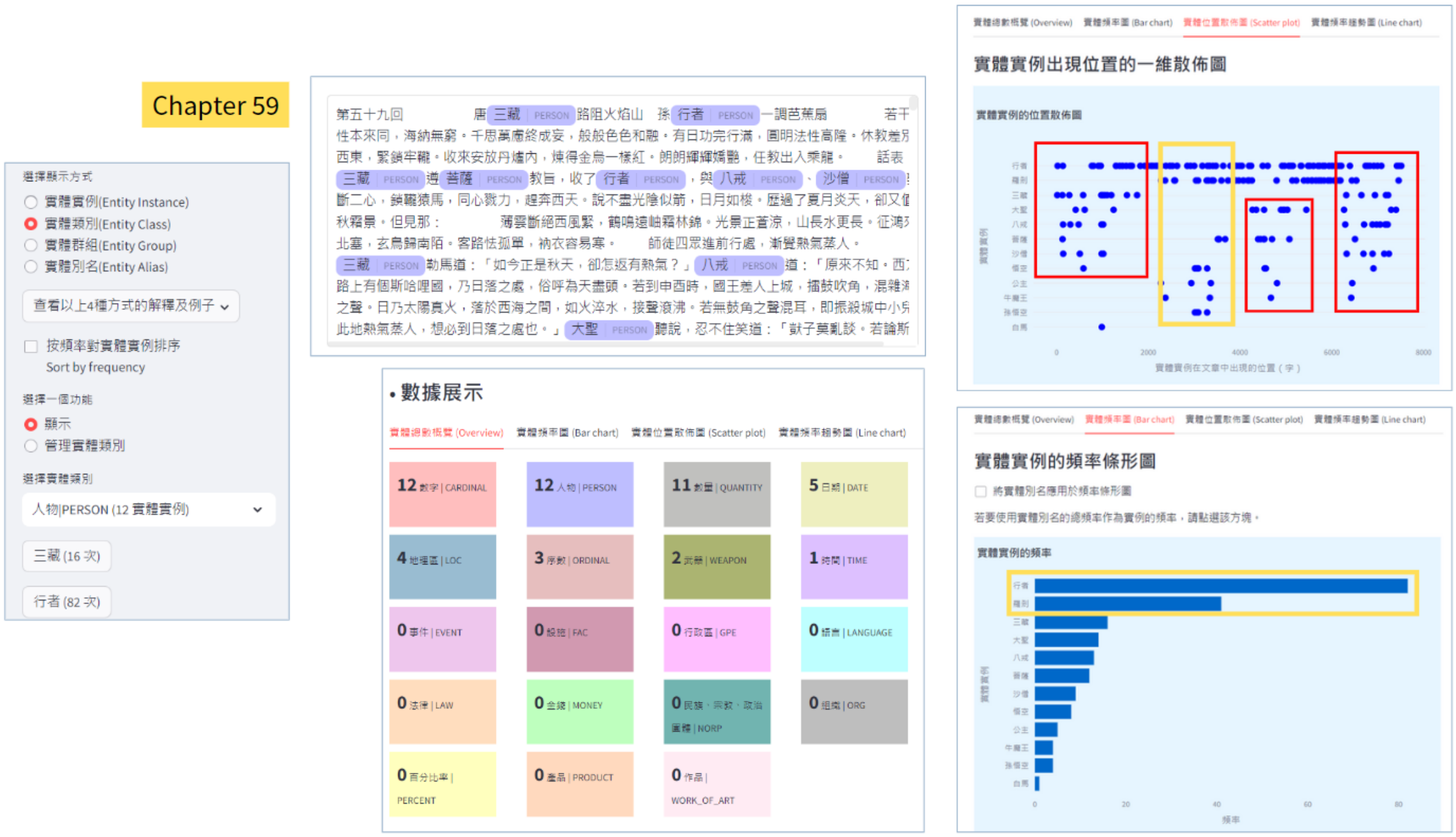

**Figure 24**

*Scatter Plot Result for Chapter 60 and 61*

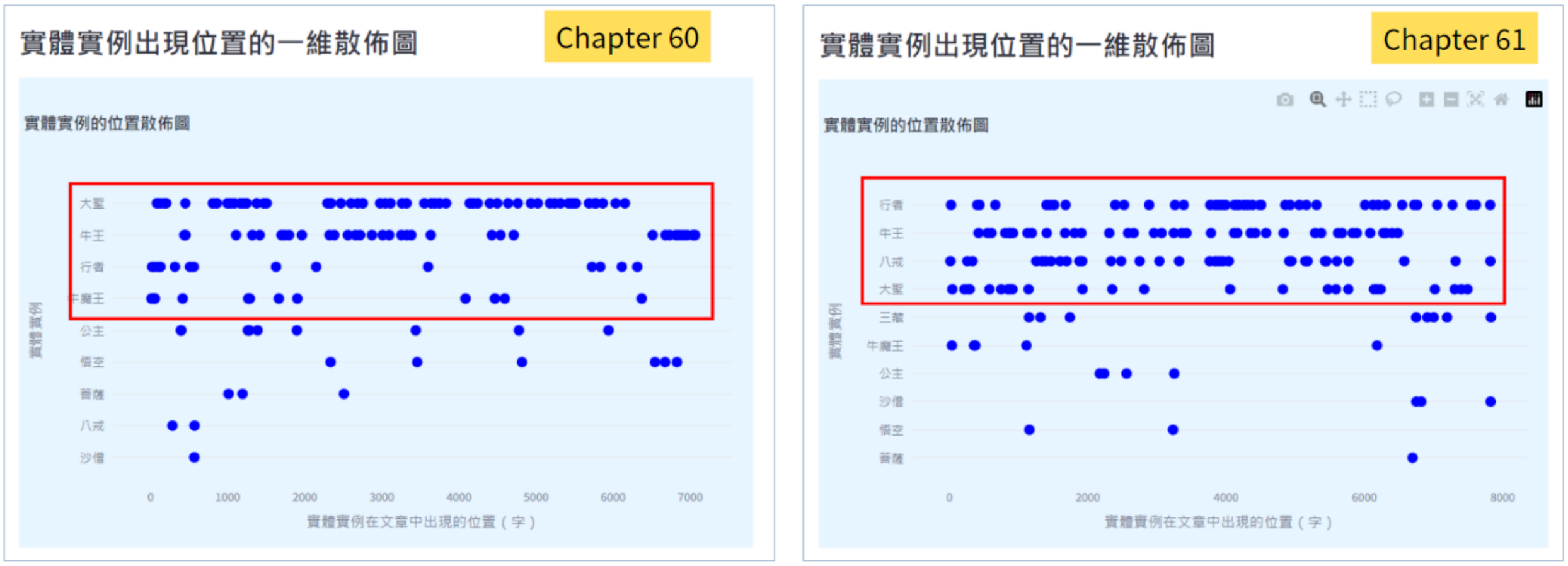

b. Entity Alias

**Figure 25**

*Example of Grouping Multiple Alternative Names into One Entity Alias for Analysis*

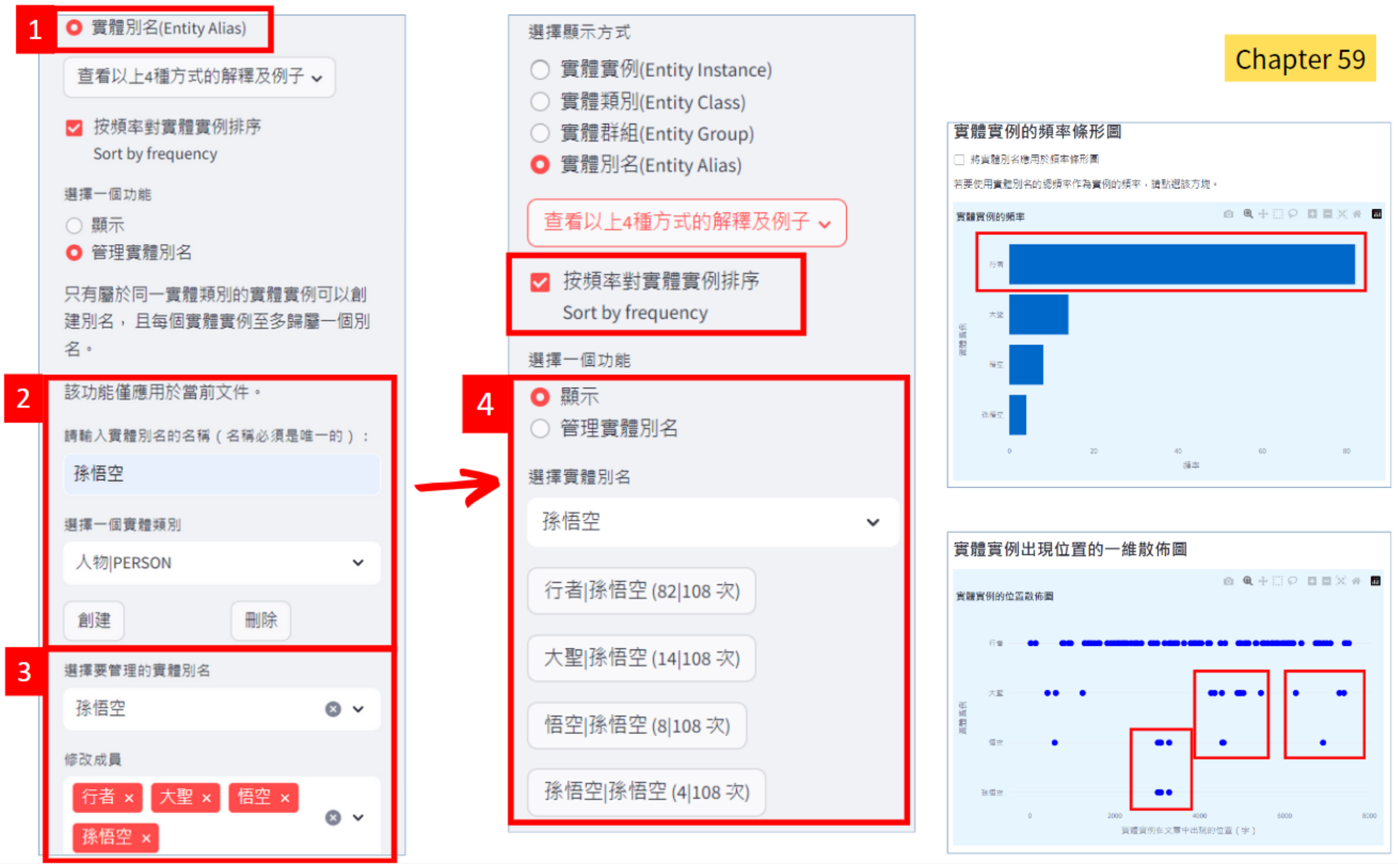

With the definition of Alias “孫悟空” above, one can study how these alternative names are used in the chapter 59. By sorting the entity list by frequency and examining the bar chart, one can know that “行者” is the major reference of “孫悟空”. From the scatter plot, one can notice that the major reference “行者” is constantly used, while other alternative names appear as clusters, which may imply that the author tend to use the same name or same set of names to refer a character in a scene (see Figure 25).

One can also define Alias “孫悟空” with the same set of entity in the other two chapters and study the overall frequency of this character across all three chapters, which is decreasing as shown in Figure 26.

**Figure 26**

*Example of using Line Chart*

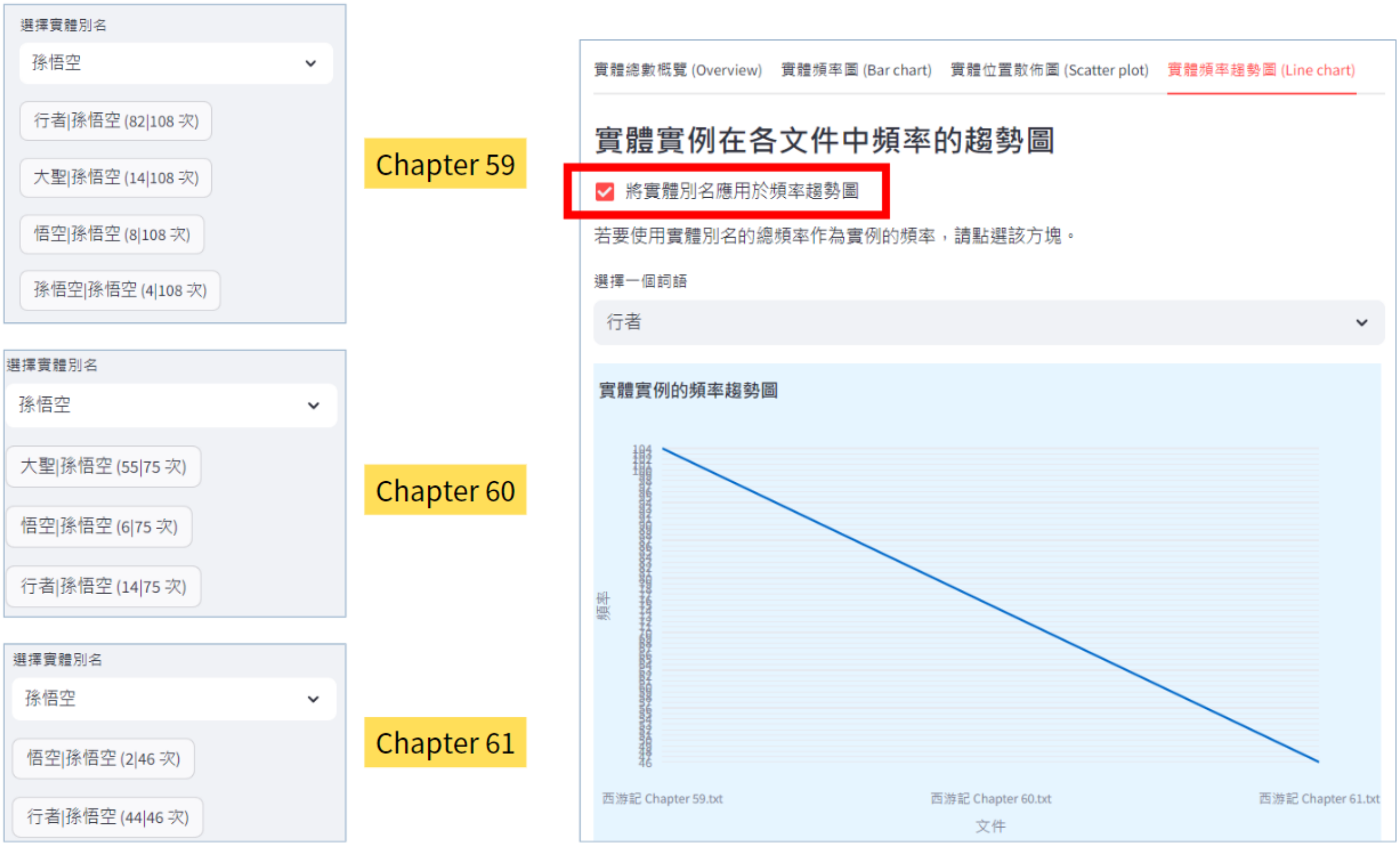

c. Entity Group

**Figure 27**

*Example of Grouping Multiple Entity Instances into One Entity Group for Analysis I*

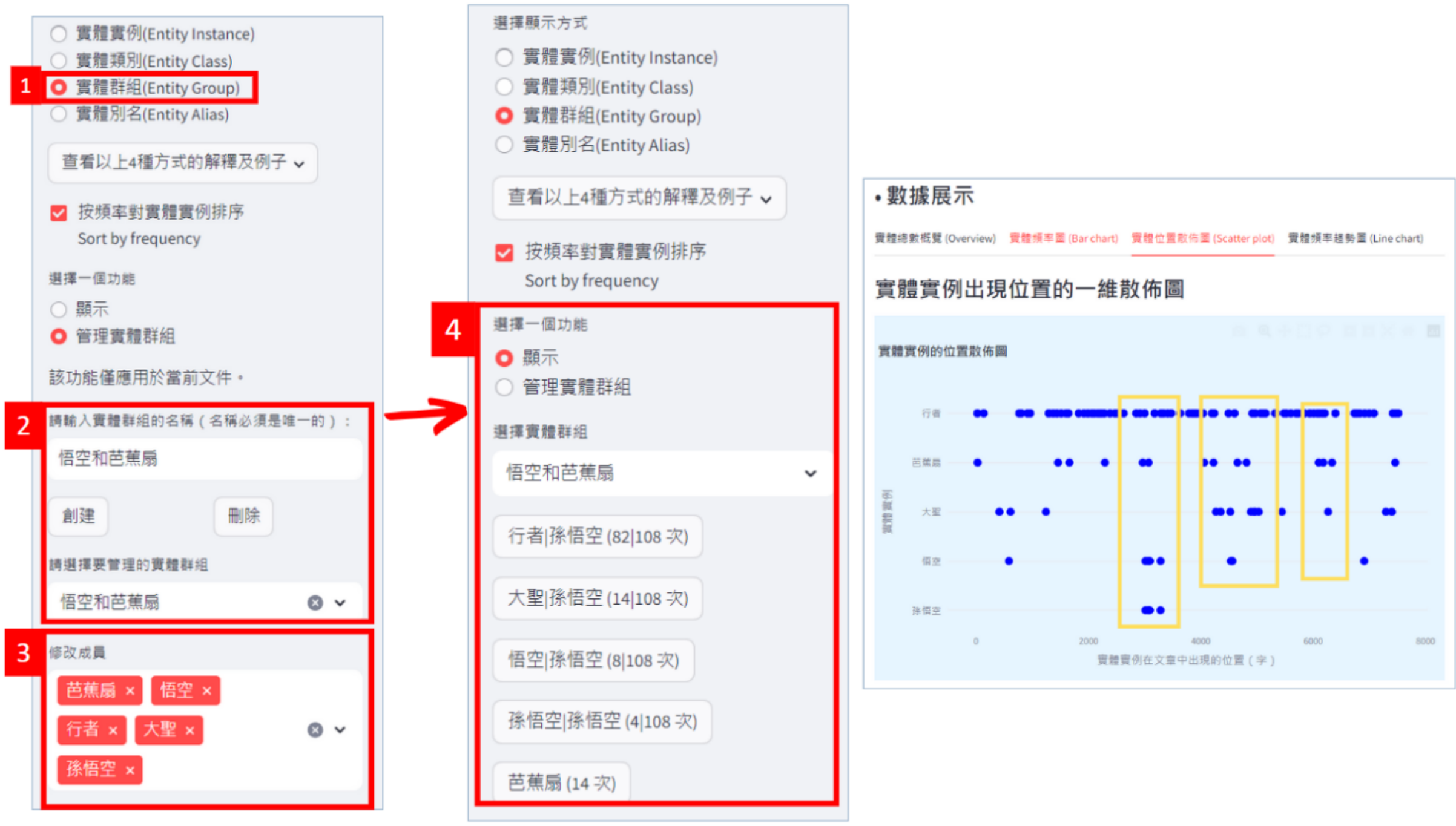

**Figure 28**

*Example of Grouping Multiple Entity Instances into One Entity Group for Analysis II*

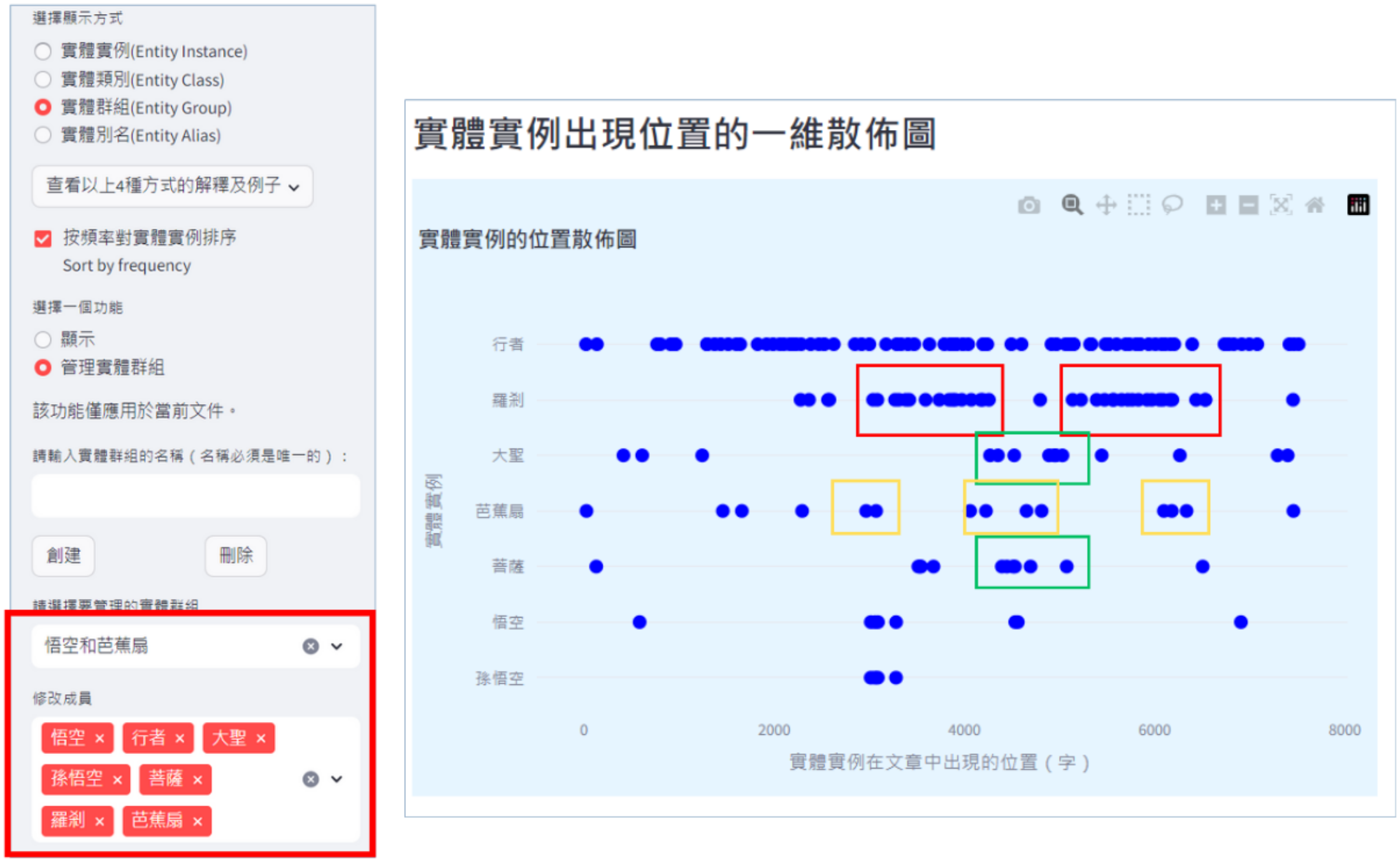

With the definition of Group “悟空和芭蕉扇” above (see Figure 27), one can filter the display with this group. In the scatter plot, there seems to be about three or more clusters of “芭蕉扇” synchronize with alternative name of “孫悟空” (except the constantly used “行者”). To get a clearer understanding, other entities can be included in the group. In this example, “羅刹” and “菩薩” are chosen. From the scatter plot, there seem to be a scene about “孫悟空” and “菩薩” between two scenes about “孫悟空” and “羅刹”, and all three scenes seems to involve “芭蕉扇” (see Figure 28). Such a rough understanding could be helpful for more detailed analysis of the text.

### 5.4.8 Data Export

After analysis, the final data about entity, group and alias can be downloaded for each file in their corresponding view page (Figure 29 and 30).

**Figure 29**

*Steps for Exporting the Entities Data*

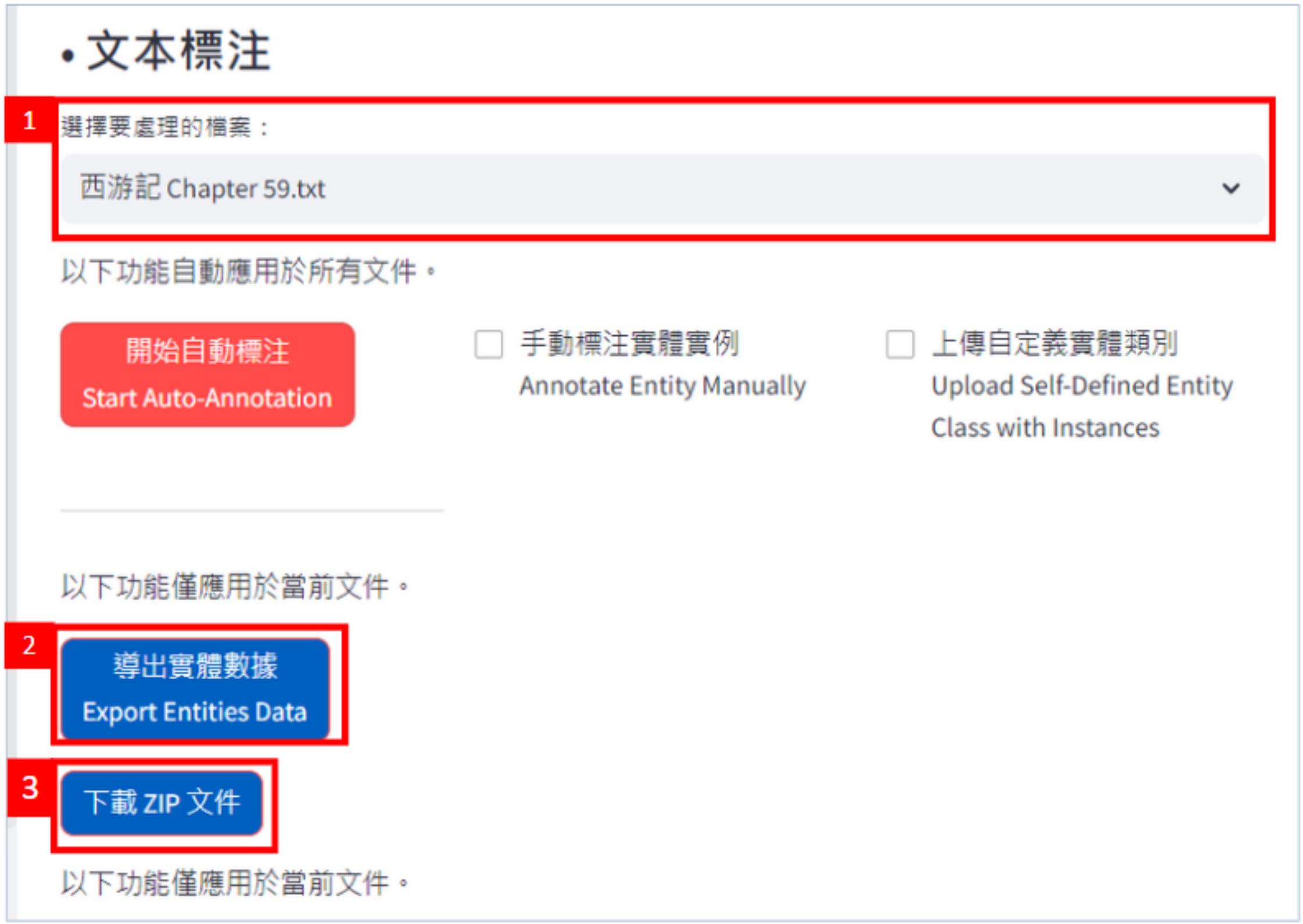

**Figure 30**

*The Exported CSV Files That Contains the Data of Entity, Alias and Group*

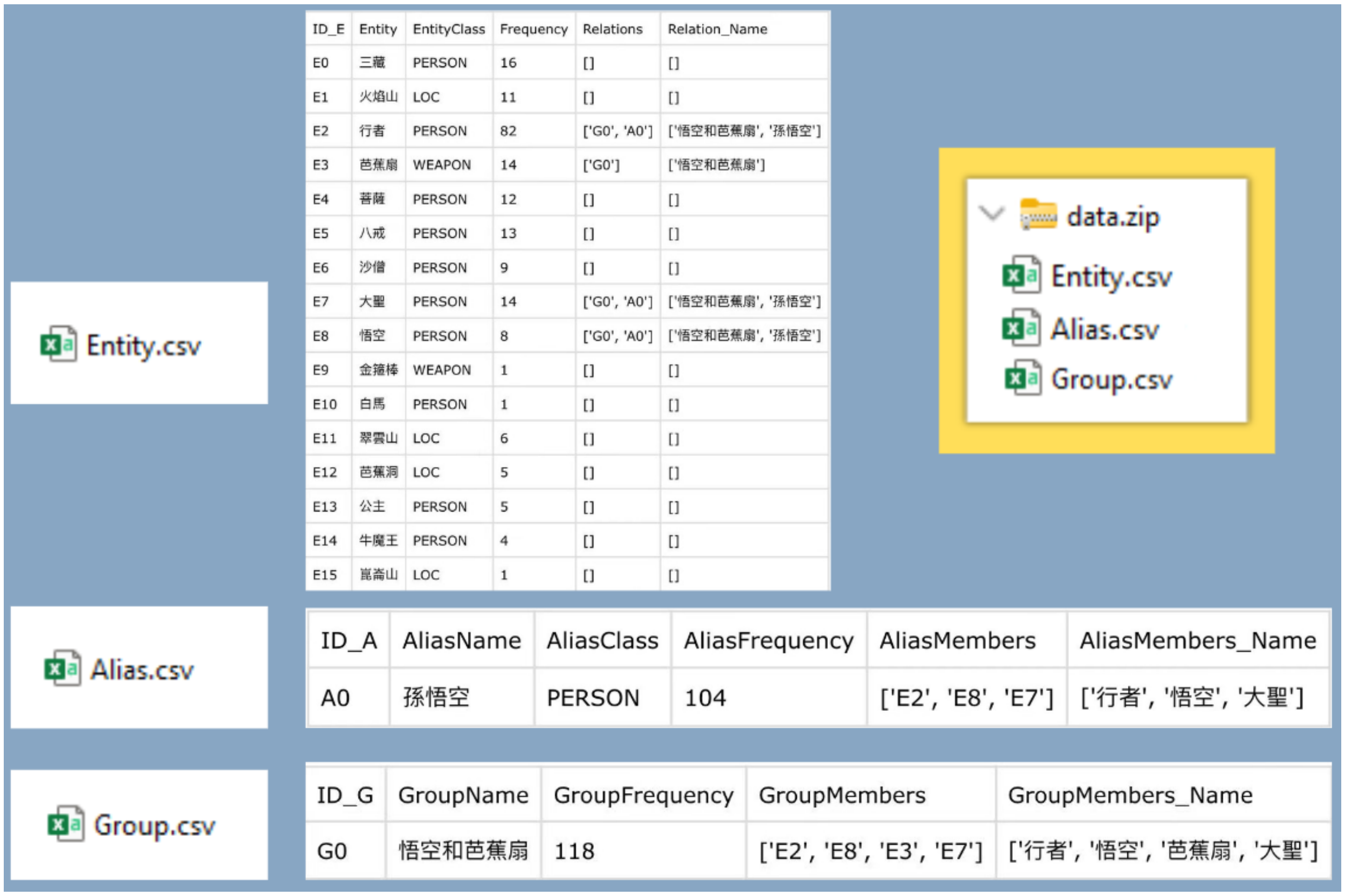

| ID_E | Entity | EntityClass | Frequency | Relations | Relation_Name |
|---|---|---|---|---|---|
| E0 | 三藏 | PERSON | 16 | [] | [] |
| E1 | 火焰山 | LOC | 11 | [] | [] |
| E2 | 行者 | PERSON | 82 | ['G0', 'A0'] | ['悟空和芭蕉扇', '孫悟空'] |
| E3 | 芭蕉扇 | WEAPON | 14 | ['G0'] | ['悟空和芭蕉扇'] |
| E4 | 菩薩 | PERSON | 12 | [] | [] |
| E5 | 八戒 | PERSON | 13 | [] | [] |
| E6 | 沙僧 | PERSON | 9 | [] | [] |
| E7 | 大聖 | PERSON | 14 | ['G0', 'A0'] | ['悟空和芭蕉扇', '孫悟空'] |
| E8 | 悟空 | PERSON | 8 | ['G0', 'A0'] | ['悟空和芭蕉扇', '孫悟空'] |
| E9 | 金箍棒 | WEAPON | 1 | [] | [] |
| E10 | 白馬 | PERSON | 1 | [] | [] |
| E11 | 翠雲山 | LOC | 6 | [] | [] |
| E12 | 芭蕉洞 | LOC | 5 | [] | [] |
| E13 | 公主 | PERSON | 5 | [] | [] |
| E14 | 牛魔王 | PERSON | 4 | [] | [] |
| E15 | 崑崙山 | LOC | 1 | [] | [] |

| ID_A | AliasName | AliasClass | AliasFrequency | AliasMembers | AliasMembers_Name |
|---|---|---|---|---|---|
| A0 | 孫悟空 | PERSON | 104 | ['E2', 'E8', 'E7'] | ['行者', '悟空', '大聖'] |

| ID_G | GroupName | GroupFrequency | GroupMembers | GroupMembers_Name |
|---|---|---|---|---|
| G0 | 悟空和芭蕉扇 | 118 | ['E2', 'E8', 'E3', 'E7'] | ['行者', '悟空', '芭蕉扇', '大聖'] |

## 6 Challenges and Obstacles

Throughout the development of this platform, we encountered numerous challenges and obstacles that highlighted the complexities of creating a robust and user-friendly tool. These issues not only pushed the technical limits of the platform but also brought attention to areas that could impact user experience and long-term functionality.

### 6.1 Solved challenges

6.1.1. Data Loss Due to Page Refreshes

**Challenge:** One of the initial challenges was the complete page refresh that occurred when buttons were clicked, leading to the loss of previously entered data. This was especially problematic for users handling large volumes of information.

**Solution:** By implementing Streamlit's session state functionality, we preserved user data across page refreshes, allowing users to retain their work even after interacting with various elements on the webpage.

6.1.2. Requiring Multiple Refreshes for Desired Outcomes

**Challenge:** Another issue was the need to refresh the page twice to display the desired data correctly. Despite debugging, the root cause could not be resolved.

**Solution:** We introduced a confirmation button that required users to click twice, ensuring accurate data display, albeit with additional user intervention.

6.1.3. Handling Multiple File Uploads

**Challenge:** When enabling the upload of multiple files, distinguishing data for each file became difficult as variables were stored in lists, complicating the separation of data between files.

**Solution:** We restructured the data handling by storing all relevant data in a dictionary format, with each file assigned a unique key. This approach streamlined the segregation of file data, allowing for smoother handling of multiple datasets.

6.1.4. File-Specific Data Display in the Sidebar

**Challenge:** When displaying entity instances from multiple files, the platform failed to determine which file was being viewed, showing all instances from all files simultaneously.

**Solution:** We divided the sidebar into tabs, corresponding to each file, enabling users to manually select which file's data to view. While this solution required user interaction, it successfully organized the data according to individual files.

### 6.2 Future improvement

6.2.1. Bulk Processing Limitations

**Limitation:** The platform is currently unable to handle the upload of large volumes of text in one go. For instance, users can upload a maximum of 10 chapters at a time, which means the entity frequency trend chart can only compare 10 chapters at once. For a book with 100 chapters, users would need to manually consolidate the trend data 10 times or export the CSV files and merge them for further analysis. This limitation hinders the convenience of large-scale literary analysis.

**Possible Future Development:** To address the bulk processing limitation, one potential solution could be to implement a more scalable file upload system. The platform could support batch uploads of larger datasets, allowing users to upload more than 10 chapters at a time. Additionally, integrating automated data merging features could help consolidate entity frequency trend data across all chapters seamlessly, eliminating the need for users to manually merge files. This improvement would greatly enhance the platform's capacity for handling large-scale literary analyses and reduce the time required for data processing.

### 6.2.2. Time-Consuming Double Confirmations

**Limitation:** To ensure data integrity and prevent accidental deletions or modifications, certain actions—such as deletions—require users to confirm their choices twice. While this adds a layer of security, it also increases the overall time needed for task completion, potentially leading to frustration for some users.

**Possible Future Development:** To reduce the time users spend on double confirmations while maintaining data security, a potential enhancement could involve offering users the option to customize their confirmation settings. For instance, a user could enable or disable double confirmation for non-critical actions or allow the platform to remember specific confirmation preferences. Another approach could involve adding a "batch confirm" feature for bulk actions, which would speed up the process by allowing users to confirm multiple changes in one go without compromising security.

### 6.2.3. Lack of Data Cloud Storage

**Limitation:** The platform currently lacks a cloud-based data storage system, meaning all data is lost if the webpage is refreshed or closed. This issue makes the tool less convenient for long-term projects or cases requiring continuous work on multiple datasets.

**Possible Future Development:** Implementing a cloud storage solution could significantly enhance the platform's usability, especially for users working on long-term projects. A future development goal could involve integrating a secure, cloud-based storage system where users can save their progress and resume their work at any time. This would allow for persistent data storage across sessions, protecting against accidental data loss due to webpage refreshes or closures. Additionally, features such as automated backups and real-time syncing could further improve user experience, ensuring that no data is lost unexpectedly.

### 6.2.4 Outdated Model Training Data

**Limitation:** The entity recognition model used in the platform was trained on data up until 2019. As a result, the model struggles to recognize new organizations, institutions, or places that have emerged in recent years. This limitation requires users to manually adjust or annotate entities, which adds to the time and effort required to process contemporary texts.

**Possible Future Development:** To improve the model's ability to recognize contemporary entities, future development could focus on updating the training data by incorporating more recent datasets that include organizations, institutions, and locations introduced after 2019. The platform could also allow for periodic model updates, ensuring that the entity recognition system stays current. Additionally, offering users the ability to upload their own training data for fine-tuning the model could provide a flexible solution, enabling the platform to better adapt to various specialized or evolving domains.

### 6.2.5. Limited Flexibility for Entity Group and Entity Alias

**Limitation:** Management of Entity Group and Alias is currently limited within one document. This might cause repetitive operation in multifile scenario, because it is possible that the same definition can be applied to several documents. For example, the same group of main characters may appear in several consecutive or even all chapters or a character with the same set of alternative names can also appear in multiple chapters, as the case in 4.4.6.

The potential of entity group and alias is not fully exploited either. In data visualization, enabling the checkboxes for "Apply Alias (將實體別名應用於……)" will only apply the total frequency of the entity alias to all member of it while retain separate representations for each in the graph. For the example in Figure 31 below, entity instance members of the alias "孫悟空" are represented separated in the bar chart, though with the same frequency. Additionally, such function is not supported in position scatter plot currently. An additional function that allows combining all members and representing them with only one abstract entity instance with the alias name might make all the visualization graphs more useful.

**Figure 31**

*Example of the Possible Future Improvement for the Bar Chart*

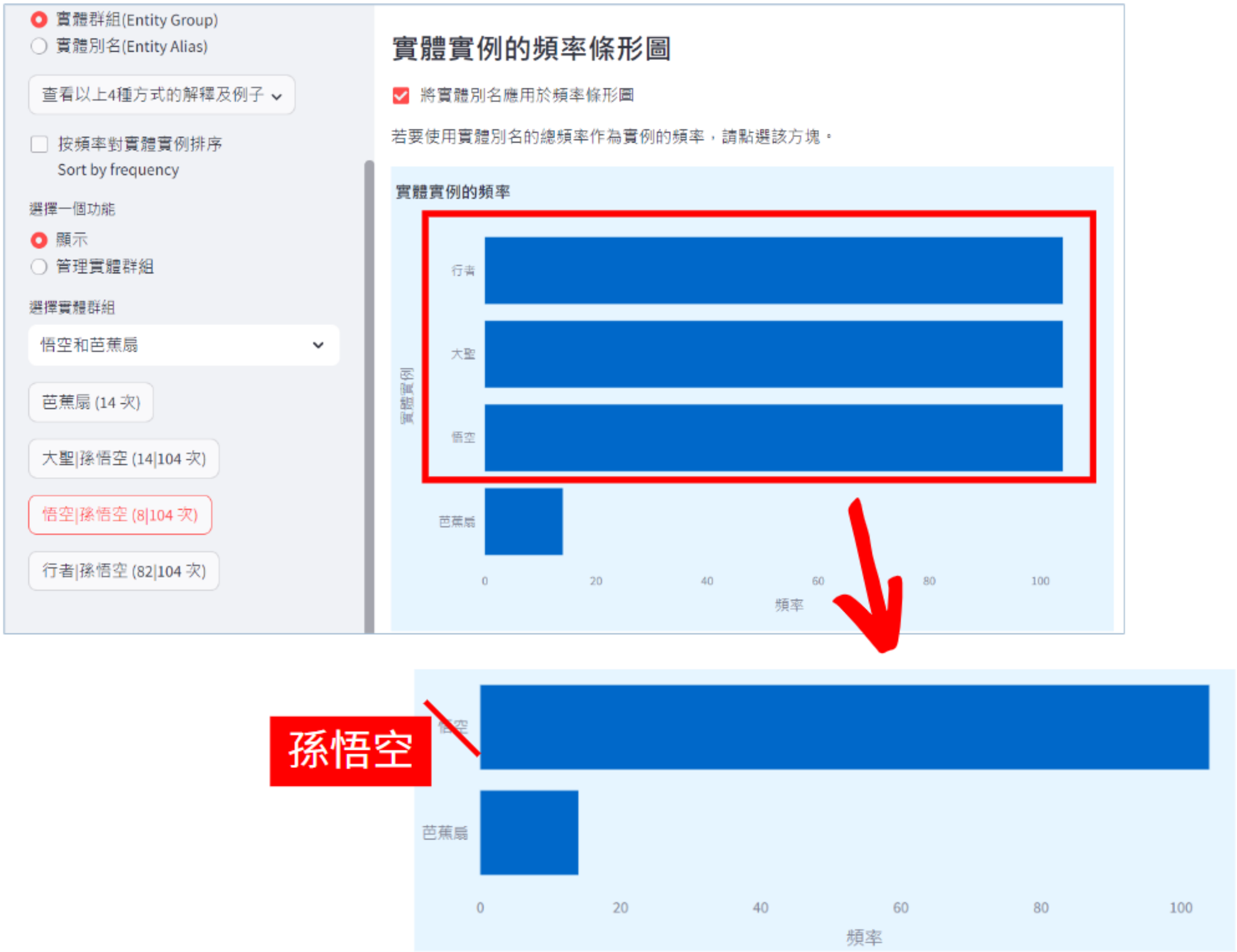

**Possible Further development:** Provide one more option for the management of Entity Group and Alias to allow applying the changes to multiple files. Specifically, it can be achieved by including an additional multi-selection box for specifying the set of documents to apply.

Through these challenges, we have identified several key areas for future improvement. Addressing these obstacles will enhance the platform's functionality, scalability, and overall user experience. Participating in this kind of project within a one-semester timeframe was particularly rewarding. We are able to be actively engaged in the entire process of tool development, from planning, functional design to programming. It pushed us to manage our time effectively, prioritize tasks, and remain adaptable in the face of unforeseen obstacles. The urgency of the timeline encouraged us to be more decisive and efficient in our work, allowing us to see tangible progress and results in a relatively short period. This practical project not only provided invaluable hands-on experience but also strengthened our teamwork and technical abilities.

## 7 Conclusion

The HKUST Library's initiative to extend AI literacy to practical utilization through the Digital Scholarship DS CoLab Projects demonstrates one of the transformative approaches to education that centers on student engagement and active learning. By enabling students to be the key contributors at every phase of the project, as demonstrated in the Chinese NER Tool development project detailed in this paper, the Library fosters an environment of experiential learning and project-based learning where theoretical knowledge is applied to real-world challenges. This hands-on experience not only cultivates critical technical skills and enhances students' problem-solving abilities, but also contributes to addressing challenges faced by the Library.

Moreover, the emphasis on collaborative learning within this framework encourages students and library staff to work together across disciplines, leveraging diverse perspectives to drive innovation. The Library's role as a catalyst for multidisciplinary collaboration reinforces the importance of teamwork in the learning process, allowing students to learn from one another while contributing to shared goals. This project exemplifies the power of student-centered pedagogy, where learners are empowered to take ownership of their education and actively shape their learning experiences. This initiative underscores the Library's commitment to nurturing a "can do spirit" among university members, illustrating how this framework prioritizes student involvement can lead to meaningful outcomes.

## Acknowledgments

We would like to thank Leo Wong (Systems & Digital Services Librarian, HKUST) for initiating Digital Scholarship in the Library, and thank Jennifer Gu (Research Support Librarian, HKUST) and Aster Zhao (Research Support Librarian, HKUST) for their support and feedback on this project.